\documentstyle[pra,aps,epsf]{revtex}
\tighten
\begin{document}
\draft
\title{Quantum channels showing superadditivity in capacity}
\author{Masahide Sasaki$^1$, Kentaro Kato$^{2}$, Masayuki Izutsu$^1$, and
 Osamu Hirota$^{2}$}
\address{${}^1$Communications Research Laboratory, Ministry of Posts and
Telecommunications\\
 Koganei, Tokyo 184, Japan}
\address{${}^2$Research Center for Quantum Communications, Tamagawa University\\
 Tamagawa-gakuen, Machida, Tokyo 194, Japan}

\date{14 October 1997}

\maketitle

\begin{abstract}
We consider a channel coding for sending classical information through a
quantum channel with a given ensemble of quantum states (letter states). 
As well known, it is generically possible in a quantum channel that the
transmittable information in block coding of length $n$
can exceed  $n$ times the maximum amount that can be sent without any
coding scheme. This so-called superadditivity in classical capacity of a
quantum channel is a distinct feature that can not be found in classical
memoryless channel.  In this paper, a practical model of channel coding
that shows this property is presented. It consists of a simple codeword
selection and the optimum decoding of the codewords minimizing the average
error probability.  At first, optimization of decoding strategy is
discussed. Then the channel coding that shows the superadditivity in
classical capacity is demonstrated. 
\end{abstract}



\section{Introduction}\label{sec1}

Theory of quantum communication was initiated more than thirty years ago,
in order to consider quantum nature of signal carrier in optical frequency
domain. In this region, one faces quite different features from RF band
communication, due to quantum noise of signal carrier itself. This theory
was then developed in 1970's revealing new aspects of information
transmission and signal detection. It now attracts much attention since new
fields such as quantum computation and cryptography emerged. Being assisted
by ideas and methods in these fields, significant progress was made in a
basic and old issue on channel capacity. In particular, the theorem was
established that the attainable maximum rate of asymptotically error free
transmission for sending classical information by using a given source of
quantum states (letter states) is precisely the Holevo bound
\cite{Holevo73_bound,Holevo79_QuantCap} (let us call it the quantum channel
coding (QCC) theorem)
\cite{Hausladen96_coding,Holevo96_coding1,Schumacher97_coding,Holevo97_coding2}. This rate is the asymptotic rate at infinite block length ,
$n\rightarrow\infty$, and is especially called {\it classical capacity} of
quantum channel. The term {\it capacity} of quantum channel is now used in
various contexts of quantum information theory, including not only
transmission of fixed classical alphabet but also sending intact quantum
states. In this paper, we confine ourselves to transmission of classical
information by use of a given letter-state ensemble, and hereafter the term
{\it capacity} is understood as {\it classical capacity} for this case.

The QCC theorem guarantees existence of codes that have the above
asymptotic property, but does not tell directly how to construct such codes
from given letter states. For practical applications, simple and systematic
coding-decoding methods at finite block length is required. Such methods
will immediately  be applied to, for example, advanced schemes of satellite
communication and ultra-fast optical fiber communication. In these cases,
signal power at receiving end might be very weak due to long distance
transmission or limited power supply so that a main source to cause error
will be nonorthogonality among letter states, which is just the situation
covered by the above  theorem.

The purpose of  this paper  is to  give some insights into practical
aspects of quantum channel coding. In quantum channel coding, block
sequences are made as direct product states of the letter states, some of
them are selected and transmitted as {\it codeword states}, and they are
then detected quantum mechanically.  A quantum channel made of the codeword
states of length $n$ is called a $n$-product channel. There are  two
essential ingredients in using this $n$-product channel; one is the
suitable selection of codeword states from all the possible sequences made
of the letter states, and the other is a {\it collective decoding} that
detects each codeword state as a single state-vector rather than decoding
the individual letter states separately. Especially, the latter fully
utilizes superposition states of the codeword states, and brings an
inseparable structure among the letter states,  which is often called {\it
entanglement}. This remarkable feature  cannot be found in classical
channel coding. 
As a consequence, the $n$-product  channel can have a {\it memory effect} 
in the sense that the channel matrix cannot be factorized into the channel
matrices corresponding to each letter, i.e., 
$P( y_1 y_2 \cdots y_n \vert x_1 x_2 \cdots x_n) 
\ne \displaystyle\prod_{i=1}^n P( y_i \vert x_i)$.  
This is even so if neither a source system emitting letter states nor a
physical process of transmission has memory effect.  This effect can be
used to increase a reliability of information transmission.  In fact, the
codeword states are selected suitably,  this effect makes it possible that
more classical information can be sent through the $n$-product channel than
$n$-times the amount that can be sent through a single use of the initial
channel made merely of the letter states without any coding scheme.    This
so-called superadditivity in  capacity is a generic nature of a quantum
channel, and is  indeed information theoretic quantum gain
\cite{Holevo73_bound,Holevo79_QuantCap}. So the first step toward finding
the ultimate channel coding for the Holevo bound might be to construct
codes that attain this quantum gain.

In this paper, quantum channels that shows the superadditivity in capacity
is described. We first consider optimization of decoding. The collective
decoding used in the proof of the QCC theorem was the so-called  {\it
square-root measurement} \cite{Hausladen96_coding}. This allows one to
derive an explicit decoding observable systematically from given codeword
states. In addition, this has been known to be almost optimum when the
quantum states to be distinguished are equally likely and almost orthogonal
\cite{Helstrom_QDET,Holevo_SubOptMeas78,Hausladen_SubOptMeas95}, which is
the case for the typical sequences obtained at very long block length.
Therefore, it played a sufficient role to evaluate the upper bound of the
decoding error.  
But this measurement is actually more than that. In Sec. \ref{sec2},  it
will be pointed out that the square-root measurement becomes precisely
optimum in terms of the average error probability in certain cases of pure
and linearly independent quantum states that are even neither equiprobable
nor almost orthogonal.  The optimality of decoding strategy should be
pursued in order to achieve performance as high as possible, especially in
a practical channel coding of finite block length. When the square-root
measurement is not optimum, there is a method to construct the optimum one
by modifying it. For a practical purpose,  we  present a basic scheme of
the optimum collective decoding of codeword states in the case of
pure-state channel in Sec. \ref{sec3}. As for physical realizations of this
scheme, the readers are referred to the subsequent paper.

In order to quantify the superadditivity in capacity, the attainable
maximum mutual information without any coding must be known.  This quantity
is usually denoted as $C_1$. The optimum solutions of the prior
probabilities and the decoding observable that maximize the mutual
information have been known only in a few cases \cite{FuchsPeres96,Ban97_C1,Osaki97_C1}.  In this paper, the most basic case of binary and pure letter
states is considered. In Sec. \ref{sec4}, we define a {\it threshold point}
where all the sequences are used as the codeword states and the accessible
information (the maximum mutual information attained by optimizing the
decoding observable  with  prior probabilities fixed) at block length $n$
is exactly $n C_1$.  At this point, the optimum decoding is not like a
collective fashion but rather reduces to the separate measurement which
detects  each letter state individually, that is, there is no room  for
generating entanglement correlation among letter states.  This threshold
point will be a useful guide for quantitative discussions. Sec. \ref{sec5}
is devoted to concluding remarks.

\section{Distinguishing linearly-independent quantum states}\label{sec2}

To begin with, we shall describe the conditions for optimality in a general
decision problem of $M$-ary quantum states.  An ensemble of quantum states
$\{\hat\rho_i\}$ is given with respective prior probabilities $\{\xi_i\}$. 
Decision process of these signal states $\{\hat\rho_i\}$  can be described
by a probability operator measure (POM) $\{\hat\Pi_i\}$ satisfying the
resolution of the identity $\displaystyle\sum_i \hat\Pi_i=\hat I$. The POM
effecting the decision needs only $M$ components $\hat\Pi_1$, $\hat\Pi_2$,
$\cdots$,  $\hat\Pi_M$ which are usually called {\it detection operators}. 
What we are seeking here are the optimum detection operators minimizing the
average error probability. 
Defining the risk operators 
$\hat W_i\equiv \xi_i \hat\rho_i$ and the Lagrange operator  
$\hat\Upsilon\equiv\displaystyle\sum_i\hat W_i\hat\Pi_i$, 
the optimum conditions  are  written as
\cite{Helstrom_QDET,Holevo73_condition},  
\begin{itemize}
\item[i)] $\hat\Pi_i(\hat W_i - \hat W_j)\hat\Pi_j =0, \quad \forall (i,j),$
\item[ii)] $\hat\Upsilon-\hat W_i \ge0, \quad \forall i.$
\end{itemize}
The minimum average error probability is given by 
\begin{equation}
P_e({\rm opt})=1-{\rm Tr}\hat\Upsilon. 
\end{equation}

When the signal states  are pure
($\hat\rho_i=\vert\rho_i\rangle\langle\rho_i\vert$ ) and linearly
independent, the optimum detection operators can be given as the
projection-valued measure (PVM) with rank 1  as 
$\hat\Pi_i=\vert\omega_i\rangle\langle\omega_i\vert$.  The set
$\{\vert\omega_i\rangle\}$ forms a complete orthonormal set in the Hilbert
space ${\cal H}_s$ spanned by 
the signal states $\{\vert\rho_i\rangle\}$ and each of them is called a
{\it measurement state}. Introducing a matrix 
${\bf X}=(X_{ij})\equiv(\langle \omega_i \vert \rho_j \rangle)$, the above
conditions are rewritten as 
\begin{itemize}
\item[i$'$)] $\xi_i X_{ii} X_{ji}^\ast = \xi_j X_{ij} X_{jj}^\ast, \quad
\forall (i,j),$
\item[ii$'$)] ${\bf T}^{(m)}\equiv(\xi_i X_{ii} X_{ji}^\ast - \xi_m X_{im}
X_{jm}^\ast) \ge0,$ 
\item[{  }] $\quad m=1, \cdots, M.$
\end{itemize} 
In general, it is a complicated job to derive explicit expressions for the
optimum measurement states satisfying the above conditions. Only in the
certain cases they have been known \cite{Osaki}. Otherwise, one has to rely
on numerical simulations like the Bayes-cost-reduction algorithm
\cite{Helstrom82}.  Most tedious part in such a method is to check the
second condition ii$'$).  But when the signal states are linearly
independent, this is ensured more simply, if 

\begin{itemize}
\item[ii$''$)] ${\bf\Upsilon}'\equiv(\xi_i X_{ii} X_{ji}^\ast)>0$ 
\end{itemize} 

\noindent
is satisfied. Its proof was given in Appendix of Ref. \cite{Helstrom82}.

Now let us consider when the square-root measurement  becomes optimum. The
square-root measurement  is defined as follows,  
\begin{mathletters} 
\begin{eqnarray}
\vert\mu_i\rangle &\equiv& \hat\rho^{-{1\over2}} \vert\tilde\rho_i\rangle, \\
\hat\rho &\equiv& \sum_{i=1}^M
\vert\tilde\rho_i\rangle\langle\tilde\rho_i\vert, \\
\vert\tilde\rho_i\rangle &\equiv& {\sqrt\xi_i}\vert\rho_i\rangle.
\end{eqnarray}
\label{eqn:sqrt-mu}
\end{mathletters}

\noindent 
As well known, the conditional probability based on  this measurement
$P(j\vert i)=\vert\langle\mu_j\vert\rho_i\rangle\vert^2$ can be calculated
by the following way. First make the Gram matrix 
${\bf\Gamma}\equiv(\langle\tilde\rho_i\vert \tilde\rho_j\rangle)$. Second
diagonalize it as, 
\begin{equation}
{\bf\Gamma}=
{\bf Q}
\left(\begin{array}{ccc} g_1  &            &   \\
                                              & \ddots &   \\
                                              &            &   g_M   
\end{array} \right)
{\bf Q}^\dagger, 
\end{equation}
where {\bf Q} is a unitary matrix. Third and finally, calculate 
\begin{equation}
{\sqrt{\bf\Gamma}}=
{\bf Q}
\left(\begin{array}{ccc} {\sqrt g_1}  &            &   \\
                                                          & \ddots &   \\
                                                          &            &  
{\sqrt {g_M}}    \end{array} \right)
{\bf Q}^\dagger.  
\end{equation}
Then the $(i,j)$-components of $\sqrt{\bf\Gamma}$ is just
$\langle\mu_i\vert\tilde
\rho_j\rangle=\langle\mu_i\vert\rho_j\rangle{\sqrt\xi_j}$. 
Here we give a useful theorem in considering the optimum collective decoding. 

\noindent
{\bf  Theorem 1}

If $\{\vert\rho_i\rangle\}$ are linearly independent, the measurement by
$\{\vert\mu_i\rangle\}$ becomes optimum when all of the diagonal components
of $\sqrt{\bf\Gamma}$ are equal. 

\noindent
{\bf Proof}

Define ${\bf Y}=(Y_{ij})\equiv (\langle\mu_i\vert\rho_j\rangle)$. The
measurement by $\{\vert\mu_i\rangle\}$ is optimum if 
\begin{itemize}
\item[i$'$)] $\xi_i Y_{ii} Y_{ji}^\ast = \xi_j Y_{ij} Y_{jj}^\ast, \quad
\forall (i,j),$
\item[ii$''$)] ${\bf\Upsilon}'\equiv(\xi_i Y_{ii} Y_{ji}^\ast)>0$,  
\end{itemize}
are satisfied. 
Denoting ${\sqrt{\bf\Gamma}}=(\tilde Y_{ij})$,  
they can be rewritten as, 
\begin{itemize}
\item[i$'$)] $\tilde Y_{ii} \tilde Y_{ji}^\ast = \tilde Y_{ij} \tilde
Y_{jj}^\ast, \quad \forall (i,j),$
\item[ii$''$)] ${\bf\Upsilon}'\equiv(\tilde Y_{ii} \tilde Y_{ji}^\ast)>0$. 
\end{itemize} 
Since ${\bf\Gamma}$ is nonnegative and Hermitian, so  is
$\sqrt{\bf\Gamma}$. Therefore the above conditions reduce to 
\begin{itemize}
\item[i$'$)] $\tilde Y_{ii} = \tilde Y_{jj}, \quad \forall (i,j),$
\item[ii$''$)] ${\bf\Upsilon}'\equiv(\tilde Y_{ii} \tilde Y_{ij})>0$. 
\end{itemize} 
Under the first condition, the second one further reduces to $\tilde
Y_{11}{\bf \tilde Y}>0$. This is automatically satisfied since  
${\sqrt{\bf\Gamma}}={\bf \tilde Y}>0$. (${\bf \tilde Y}>0$ implies $\tilde
Y_{ii}>0$, $\forall i$). Thus for the square-root measurement, the first
condition just above is enough for the optimum condition and this means the
theorem. \hfill$\Box$ 

Thus the square-root measurement plays a practical role not only in the
case of equally probable and almost orthogonal states but also the case
that the signal states satisfy the above condition.  The related discussion
was given by Ban et. al.  in the case of equally probable and symmetric
states \cite{BanKurokawa_SqRt}.  
Even if the signal states do not satisfy the above condition,
$\{\vert\mu_i\rangle\}$ can be  good initial states in searching the
optimum measurement states. At first,  note that the following remark. 

\noindent
{\bf Remark}

If $\{\vert\rho_i\rangle\}$ are linearly independent,
$\{\vert\mu_i\rangle\}$ are orthonormal.

\noindent
{\bf Proof}

The optimum measurement states $\{\vert\omega_i\rangle\}$ is a complete
orthonormal set in ${\cal H}_s$. 
Define $\hat X\equiv\displaystyle\sum_{i,j}X_{ij}
\vert\omega_i\rangle\langle\omega_j\vert$ so that $\vert\rho_i\rangle=\hat
X\vert\omega_i\rangle$. Because of the linearly independence of
$\{\vert\rho_i\rangle\}$, $\hat X$ is nonsingular and $\hat X^{-1}$ exists.
Then 
\begin{eqnarray*}
\langle\mu_i\vert\mu_j\rangle
&=&\langle\tilde\rho_i\vert  \hat \rho^{-1} \vert\tilde\rho_j\rangle  \\
&=&{\sqrt\xi_i}\langle\omega_i\vert  \hat X^\dagger\hat \rho^{-1} \hat
X\vert\omega_j\rangle{\sqrt\xi_j}  \\
&=&{\sqrt\xi_i}\langle\omega_i\vert  \left( \sum_k \xi_k \hat X^{-1}
\vert\rho_k\rangle\langle\rho_k\vert \hat X^{\dagger -1}\right)^{-1}
\vert\omega_j\rangle{\sqrt\xi_j}  \\
&=&{\sqrt\xi_i}\langle\omega_i\vert  \left( \sum_k \xi_k
\vert\omega_k\rangle\langle\omega_k\vert \right)^{-1}
\vert\omega_j\rangle{\sqrt\xi_j}  \\
&=&\delta_{ij}. 
\end{eqnarray*}
\hfill$\Box$

\noindent
Thus, for linearly independent states, the set $\{\vert\mu_i\rangle\}$ is
always a complete orthogonal set in ${\cal H}_s$. So it can be connected
via a unitary operator $\hat U$ in ${\cal H}_s$ with the optimum
measurement states $\{\vert\omega_i\rangle\}$ as $\vert\omega_i\rangle=\hat
V\vert\mu_i\rangle$.  Such an operator can be constructed, for example, as
a series of 2-dim rotations by applying the Bayes-cost-reduction algorithm
\cite{Helstrom82}.  This algorithm consists of steps of solving a binary
decision problem of a chosen pair of signal states $\{ \vert\rho_i\rangle ,
\vert\rho_j\rangle \}$ on the plane spanned by the corresponding pair of
basis vectors $\{ \vert\mu_i\rangle , \vert\mu_j\rangle \}$. At every step,
two basis vectors are revised, and the average error would decrease or, at
worst, remain the same. These 2-dim rotations are continued till reaching
the optimum point where the previous conditions i$'$) and ii$'$) are
satisfied. The resulting series of their products is just the required
unitary operator $\hat V$.

\section{Optimum collective decoding of codewords}\label{sec3}

Deriving the analytic expression of the optimum measurement basis vectors
$\{\vert\omega_i\rangle\}$ is a difficult job, but these basis vectors can
be constructed somehow as explained in the previous section, and at the
same time the channel matrix can be obtained. Thus only for evaluating
performance, it is  sufficient to derive these $\{\vert\omega_i\rangle\}$.
From a  practice point of view, however, the basis vectors
$\{\vert\omega_i\rangle\}$ hardly imply a corresponding physical process.
Although the set $\{\vert\omega_i\rangle\}$ forms a standard von Neumann
measurement, its physical implementation usually remains a nontrivial 
problem.  In this section, we present a useful scheme for  realizing the
optimum collective decoding. In the case that the letter states are binary,
this scheme naturally leads to an implementation based on a quantum circuit
and a well-defined  physical measurement.

Let  binary letter states be $\{ \vert+\rangle , \vert-\rangle\}$ whose
state overlap $\kappa=\langle+\vert-\rangle$ is assumed to be real and to
lie in $0\le\kappa<1$.  By $n$-th extension, we pick up $M$-ary codeword 
states $\{  \vert S_1\rangle, \cdots,   \vert S_M\rangle   \}$ $(M\le2^n)$
from the $2^n$ possible sequences of length $n$, and use them with
respective input probabilities $\{ \zeta_1, \cdots, \zeta_M \}$.  The rest
of sequences are denoted as $\{  \vert S_{M+1}\rangle, \cdots,   \vert
S_{2^n}\rangle   \}$.  Since the codeword states are linearly independent,
they span the $M$-dim Hilbert space ${\cal H}_s$.  Let the $n$-th extended
Hilbert space be ${\cal H}_s^{\otimes n}$. The optimum collective decoding
is described by the orthonormal basis vectors $\{  \vert\omega_1\rangle,
\cdots,   \vert\omega_M\rangle   \}$ derived in such a way as mentioned in
the previous section.        
An orthonormal set $\{  \vert\omega_1\rangle, \cdots,  
\vert\omega_{2^n}\rangle   \}$ in the extended space ${\cal H}_s^{\otimes
n}$ can be made by adding the other basis vectors obtained by using the
Schmidt orthogonalization, 
\begin{equation}
\vert\omega_i\rangle=
{
{\vert
S_i\rangle-\displaystyle\sum_{k=1}^{i-1}\vert\omega_k\rangle\langle\omega_k\
vert S_i\rangle}
\over
{\sqrt{1-\displaystyle\sum_{k=1}^{i-1}\vert\langle\omega_k\vert
S_i\rangle\vert^2}}  
 },  \quad (i=M+1,\cdots, 2^n). 
\label{eqn:orthogonal_set}
\end{equation}
We denote the expansion of all the sequences by the above basis vectors as,  
\begin{mathletters} 
\begin{equation}
\left(\begin{array}{c} \vert S_1\rangle \\
                                   \vdots \\
                                   \vert S_{2^n}\rangle    \end{array} \right)
=
{\bf B}
\left(\begin{array}{c} \vert \omega_1\rangle \\
                                   \vdots \\
                                   \vert \omega_{2^n}\rangle    \end{array}
\right),  
\end{equation}
\begin{equation}
{\bf B}=(B_{ij})=(\langle \omega_j\vert S_i\rangle).
\end{equation}
\end{mathletters} 

\noindent    
Making two orthonormal basis vectors $\{ \vert a\rangle , \vert b\rangle\}$
from $\{ \vert+\rangle , \vert-\rangle\}$, we introduce the $2^n$ product
basis vectors, 
\begin{equation}
\begin{array}{ccl}
\vert A_1\rangle&\equiv&\vert a\rangle\otimes\vert
a\rangle\otimes\cdots\otimes\vert a\rangle\otimes\vert a\rangle, \\
\vert A_2\rangle&\equiv&\vert a\rangle\otimes\vert
a\rangle\otimes\cdots\otimes\vert a\rangle\otimes\vert b\rangle, \\
                                                                             & \vdots & \\
\vert A_{2^n-1}\rangle&\equiv&\vert b\rangle\otimes\vert
b\rangle\otimes\cdots\otimes\vert b\rangle\otimes\vert a\rangle, \\
\vert A_{2^n}\rangle&\equiv&\vert b\rangle\otimes\vert
b\rangle\otimes\cdots\otimes\vert b\rangle\otimes\vert b\rangle,
\end{array}
\label{eqn:allbasis}
\end{equation}
and denote another expansion by them as, 
\begin{mathletters} 
\begin{equation}
\left(\begin{array}{c} \vert S_1\rangle \\
                                   \vdots \\
                                   \vert S_{2^n}\rangle    \end{array} \right)
={\bf C}  \left(\begin{array}{c} \vert A_1\rangle \\
                                   \vdots \\
                                   \vert A_{2^n}\rangle    \end{array}
\right),   
\label{eqn:expansion_C} 
\end{equation}
\begin{equation}
{\bf C}=(C_{ij})=(\langle A_j\vert S_i\rangle).
\label{eqn:matrix_C}
\end{equation}
\end{mathletters} 

\noindent  
The two basis sets are connected via a unitary operator $\hat U$ on ${\cal
H}_s^{\otimes n}$ as,
\begin{mathletters}
\begin{equation}
\vert \omega_i\rangle =\hat U^\dagger \vert A_i\rangle, \quad (i=1,\cdots,2^n),
\end{equation}
where 
\begin{equation}
\hat U^\dagger = \sum_{i,j}^{2^n} u_{ji} \vert A_j\rangle\langle A_i \vert,
\quad u_{ji}=({\bf B}^{-1} {\bf C})_{ij}. 
\end{equation}
\label{eqn:def_U}
\end{mathletters}

\noindent  
Here the optimum collective decoding can be described by the set $\{ \hat
U^\dagger \vert A_1\rangle, \cdots,    \hat U^\dagger \vert A_M\rangle \}$
The minimum error probability is obtained as 
\begin{equation}
P_e({\rm opt})=1-\sum_{m=1}^M \zeta_m \vert\langle S_m\vert \hat
U^\dagger\vert A_m\rangle\vert^2. 
\end{equation}
This clearly means that the optimum collective decoding
$\{\vert\omega_m\rangle\}$ can be effected by (i) transforming the codeword
states $\{\vert S_m\rangle\}$ by the unitary transformation $\hat U$, and
(ii) applying the von Neumann measurement $\{ \vert A_m\rangle\langle
A_m\vert\}$ into the transformed codeword states. This type of detection
scheme is called the received quantum state control
\cite{Hirota88_RQSC,Sasaki's}.  The final measurement is  actually a
separate measurement distinguishing each output letter state as $\vert
a\rangle$ or $\vert b\rangle$ sequentially.

Now the measurement basis vectors need not to be the above combination but
may be chosen as any combination of $M$ distinct elements of product basis
vectors $\{\vert A_{i_1}\rangle, \cdots, \vert A_{i_M}\rangle\}$. Depending
on  the choice, the matrix $\bf C$ should be redefined by rearranging the
order of elements of the vectors of the right-hand side in Eq.
(\ref{eqn:expansion_C}), as 
$\{\vert A_{i_1}\rangle, \cdots, \vert A_{i_M}\rangle , \vert
A_{i_{M+1}}\rangle, \cdots,  \vert A_{i_{2^n}}\rangle \}$ .  Then the
unitary operator $\hat U$ constructed by Eq. (\ref{eqn:def_U})  transforms
the codeword states adaptively to the chosen basis vectors $\{\vert
A_m\rangle\}$ such that the minimum average error probability is attained
by the separate measurement. Note that $\hat U$ acts on the $2^n$-dim
Hilbert space ${\cal H}_s^{\otimes n}$ rather than the $M$-dim space ${\cal
H}_s$. After the unitary transformation has been carried out, the resulting
sequences at the final measurement always lie in the space spanned by 
$\{\vert A_{i_1}\rangle, \cdots, \vert A_{i_M}\rangle\}$.  If the
transformation is skipped, all the product basis vectors will come out. 
The channel model of this scheme is illustrated in the case of $n=3$ and
$M=4$.

This kind of decomposition makes it easier to design the collective
decoding systematically. As the final measurement on each letter state
system, each process of which is described by the set $\{\vert a\rangle ,
\vert b\rangle\}$, the most suitable and implementable method  may be
chosen. Main problem is the realization of the unitary transformation as an
{\it adaptor}  to the final measurement. Corresponding physical processes
are sometimes subtle.  Difficulty of finding them  may be case by case
depending on what kind of letter-state system is provided. However if a 
2-bit gate  acting on qubits made of the two basis vectors $\{\vert
a\rangle , \vert b\rangle\}$ is available, the required unitary
transformation can be, in principle, effected as a quantum circuit used in
quantum computation.

Barenco. et. al.  already showed that an exact {\it simulation} of any
discrete unitary operator can be carried out by using a quantum computing
network \cite{Barenco95}. This story can be directly translated into the
real operation of $\hat U$ on the codeword states.  
At first, $\hat U$ is decomposed into  $U(2)$-operators $\hat T_{[j,i]}$ by
applying the algorithm proposed by Reck and others \cite{Reck94} as, 
\begin{mathletters}
\begin{equation}
\hat U =\hat T_{[2,1]}  \hat T_{[3,1]} \cdots \hat T_{[2^n,2^n-2]} \hat
T_{[2^n,2^n-1]}, \label{eqn:decomp_a} 
\end{equation}
where 
\begin{equation}
\hat T_{[j,i]} ={\rm exp}[-\gamma _{ji}(\vert A_i\rangle\langle A_j\vert -
\vert A_j\rangle\langle A_i\vert)]       . \label{eqn:decomp_b}
\end{equation}
\end{mathletters}

\noindent
Then the above 2-dim rotations $\hat T_{[j,i]}$'s are converted into
quantum circuits  by using the formula established by Barenco et. al.
\cite{Barenco95}.  
Here the following  point should be noted. Since qubits are letter states
themselves constituting the codeword states, the gates should consist of
the single physical species from which the letter states are made. Such
gates known so far are Sleator and Weinfurter's gate consisting of
two-state atoms \cite{Sleator95} and the quantum phase gate acting on two
photon-polarization states \cite{Turchette95}.  In the subsequent paper,
examples of the required quantum circuits are described based on such
gates.

\section{superadditivity in capacity of quantum channel}\label{sec4}

The purpose of this section is to demonstrate simple codes that  show the
superadditivity in  capacity.     Let us introduce some definitions. 
Prepare an ensemble of letter states $\hat{\bf s}=\{ \hat s_1, \cdots, \hat
s_l  \}$ in a Hilbert space ${\cal H}_s$. They represent input letters $\{
1, \cdots, l  \}$.  Let 
$\mbox{\boldmath $\xi$}=\{ \xi_1, \cdots, \xi_l  \}$ be corresponding prior
probabilities. 
A decoding process is described by a probability operator measure (POM) on
${\cal H}_s$, 
$\hat{\mbox{\boldmath $\pi$}}=\{ \hat\pi_1, \cdots, \hat\pi_{l'}  \}$
representing output letters $\{ 1, \cdots, l'  \}$. We call the mapping $\{
1, \cdots, l  \}\mapsto\{ 1, \cdots, l'  \}$ the initial quantum channel. 
Fixing $\mbox{\boldmath $\xi$}$, $\hat{\bf s}$ and $\hat{\mbox{\boldmath
$\pi$}}$,  the mutual information is defined as 
\begin{equation}
I_1(\mbox{\boldmath $\xi$},\hat{\bf s}:\hat{\mbox{\boldmath $\pi$}})
=\sum_{i=1}^l \xi_i \sum_{j=1}^{l'} p(j|i) \log \frac{p(j|i)}
{\displaystyle\sum_{k=1}^l \xi_k p(j|k)},
\label{eqn:mutual_info_1}
\end{equation}
where $p(j|i)=\text{Tr}(\hat\pi_j \hat s_i)$ is a conditional probability
that the letter $j$ is chosen when the letter $i$ has been sent.  The
maximum value of this quantity optimized with respect to  $\mbox{\boldmath
$\xi$}$ and $\hat{\mbox{\boldmath $\pi$}}$ is usually denoted as $C_1$, 
\begin{equation}
C_1(\hat{\bf s})\equiv 
\sup_{ \mbox{\boldmath $\xi$} ,\hat{\mbox{\boldmath $\pi$}}}
I_1(\mbox{\boldmath $\xi$},\hat{\bf s}:\hat{\mbox{\boldmath $\pi$}}). 
\label{eqn:C_1_def}
\end{equation}

A basic channel coding consists of (i) concatenation of the letter states
into the $l^n$ block sequences 
$\{ \hat s_{i_1}\otimes\cdots\otimes\hat s_{i_n}\}$, (ii) pruning of them
into $M$-ary codewords  $\hat{\bf S}=\{ \hat S_1, \cdots, \hat S_M  \}$
which can encode $\log_2 M$ classical bits, and (iii) finding an
appropriate decoding POM $\hat{\bf\Pi}=\{ \hat\Pi_1, \cdots, \hat\Pi_{M'} 
\}$ in the 
extended Hilbert space ${\cal H}_l^{\otimes n}$. The obtained channel is
called the $n$-th extended channel. Assigning input distribution
$\mbox{\boldmath $\zeta$}=\{ \zeta_1, \cdots, \zeta_M  \}$ to the
codewords,  the mutual information is defined also for this channel as, 
\begin{equation}
I_n(\mbox{\boldmath $\zeta$},\hat{\bf S}:\hat{\bf\Pi})
=\sum_{i=1}^M \zeta_i \sum_{j=1}^{M'} P(j|i) \log \frac{P(j|i)}
{\displaystyle\sum_{k=M}^l \zeta_k P(j|k)},
\label{eqn:mutual_info_n}
\end{equation}
where $P(j|i)=\text{Tr}(\hat\Pi_j \hat S_i)$. Let us define the $n$-{\it th
order capacity} as, 
\begin{equation}
C_n(\hat{\bf s})\equiv \sup_{\mbox{\boldmath $\zeta$},\hat{\bf S},\hat{\bf\Pi}}
I_n(\mbox{\boldmath $\zeta$},\hat{\bf S}:\hat{\bf\Pi}). 
\label{eqn:C_n_def}
\end{equation}
Then generally, $C_n(\hat{\bf s})\ge n C_1(\hat{\bf s})$ holds for a
quantum channel \cite{Holevo73_bound,Holevo79_QuantCap}.  
This property is the superadditivity in capacity. One can define the limit
$C(\hat{\bf s})\equiv\displaystyle\lim_{n\rightarrow\infty}C_n(\hat{\bf
s})/n$. 
The quantum channel coding theorem 
\cite{Hausladen96_coding,Holevo96_coding1,Schumacher97_coding,Holevo97_coding2}  says that this $C(\hat{\bf s})$ is just the attainable rate of
asymptotically error free transmission, hence the {\it intrinsic capacity}
of the initial quantum channel, and is exactly equal to the Holevo bound, 
\begin{equation}
C(\hat{\bf s})=\sup_{\mbox{\boldmath $\xi$}} 
\left[ H(\displaystyle\sum_i\xi_i\hat s_i)   
                                          -   \displaystyle\sum_i  \xi_i
H(\hat s_i) \right],
\label{eqn:Holevo_bound}
\end{equation}
where $H(\hat s_i)\equiv-{\rm Tr}(\hat s_i\log\hat s_i)$ is the von Neumann
entropy of the density operator $\hat s_i$.  This theorem ensures that
there exist such codes that the decoding error vanishes asymptotically as
$n\rightarrow\infty$ if the transmission rate $R={1\over n}\log_2 M$ is
kept below $C(\hat{\bf s})$.

A remaining big problem is to find such codes. For this purpose, the first
thing to be understood is the superadditivity in capacity. It should be
stressed that the strict superadditivity $C_n(\hat{\bf s})>n C_1(\hat{\bf
s})$ is definitely impossible in classical memoryless channel. In contrast,
a quantum channel has a {\it memory effect} seen in channel matrix as 
$P( y_1 y_2 \cdots y_n \vert x_1 x_2 \cdots x_n) 
\ne \displaystyle\prod_{i=1}^n P( y_i \vert x_i)$, 
even when the source of letter states and the physical transmission channel
do not have any memory effects. This {\it memory effect} is caused by the
decoding process itself. That is, when a {\it collective decoding} is
applied to the codewords, the entanglement structure among the letter
states prevents, in general, the channel matrix from being factorized as
the above. 
For attaining the strict superadditivity, an appropriate {\it memory
effect} need to be generated by an appropriate selection of codewords {\it
and} a collective decoding for them. This property is thus indeed quantum
gain in information transmission.

The essential role of entanglement for  the information theoretic quantum
gain can be stressed rigorously by the following theorem. 

\noindent
{\bf Theorem 2}

Suppose that two ensembles of letter states 
$\hat{\bf s}^{(1)}=\{\hat s_i^{(1)}\}$ in ${\cal H}_s^{(1)}$ and 
$\hat{\bf s}^{(2)}=\{\hat s_j^{(2)}\}$ in ${\cal H}_s^{(2)}$ are given, 
and the first order capacities $C_1(\hat{\bf s}^{(1)})$ and $C_1(\hat{\bf
s}^{(2)})$ are attained for the prior probabilities 
${\mbox{\boldmath $\xi$}}^{(1)}$ and ${\mbox{\boldmath $\xi$}}^{(2)}$, and
the detection operators $\hat{\mbox{\boldmath $\pi$}}^{(1)}$ and
$\hat{\mbox{\boldmath $\pi$}}^{(2)}$, respectively. Then the accessible
information of the channel with the inputs 
$\hat{\bf s}^{(1)}\otimes\hat{\bf s}^{(2)}$ and the prior probabilities 
${\mbox{\boldmath $\xi$}}^{(1)}\otimes{\mbox{\boldmath $\xi$}}^{(2)}$ is
given as 
\begin{equation}
\sup_{\hat{\bf\Pi}} 
I(\mbox{\boldmath $\xi$}^{(1)}\otimes\mbox{\boldmath $\xi$}^{(2)},
    \hat{\bf s}^{(1)}\otimes\hat{\bf s}^{(2)}:\hat{\bf\Pi})
    =C_1(\hat{\bf s}^{(1)})+C_1(\hat{\bf s}^{(2)}), 
\end{equation}
when 
$$
\hat{\bf\Pi}=
\hat{\mbox{\boldmath $\pi$}}^{(1)}\otimes\hat{\mbox{\boldmath $\pi$}}^{(2)}. 
$$ 
The proof is given in Appendix A \cite{Holevo97_private_commun}.

Now suppose that the supremum in Eq. (\ref{eqn:C_1_def}) is attained when
$\mbox{\boldmath $\xi$}=\mbox{\boldmath $\xi$}^\ast$ and 
$\hat{\mbox{\boldmath $\pi$}}=\hat{\mbox{\boldmath $\pi$}}^\ast$. If all of
the $l^n$ sequences 
$\{ \hat s_{i_1}\otimes\cdots\otimes\hat s_{i_n}\}$ are used as the
codewords with fixed prior probabilities 
$\{ \xi_{i_1}^\ast \times \cdots \times   \xi_{i_n}^\ast \}$, according to
the above theorem, the optimum decoding $\hat{\bf\Pi}$ maximizing the
mutual information 
$I_n(\mbox{\boldmath $\xi$}^{\ast\otimes n}, \hat{\bf s}^{\otimes
n},\hat{\bf\Pi})$ is 
\begin{equation}
\hat\Pi_{i_1\cdots i_n}=\hat\pi_{i_1}^\ast \otimes \cdots \otimes 
\hat\pi_{i_1}^\ast.
\label{eqn:threshold_pi_inf}
\end{equation}
The accessible information is simply $n$ times $C_1(\hat{\bf s})$, 
\begin{equation}
\sup_{\hat{\bf\Pi}} 
I_n(\mbox{\boldmath $\xi$}^{\ast\otimes n}, \hat{\bf s}^{\otimes
n},\hat{\bf\Pi})
    =n C_1(\hat{\bf s}). 
\label{eqn:threshold_I}
\end{equation} 
Thus in this restricted case, there is no room for entanglement to be
generated.  So this case provides  a {\it threshold point} for the
information theoretic quantum gain.  Once the input probabilities are
redistributed so as to reduce weights of some codewords,  the entanglement
becomes possible, and the quantum gain can be obtained by an appropriate
decoding $\hat{\bf\Pi}$.  
Concerning to the threshold point,  it might be worth mentioning a similar
theorem in terms of the average error probability.  

\noindent
{\bf Theorem 3}

For given states  $\{\hat s_i\}$ with prior probabilities  $\{\xi_i\}$, let
$\{\hat\pi_i\}$ be the optimum POM minimizing the average error
probability. Then in distinguishing the product states 
$\{ \hat s_{i_1}\otimes\cdots\otimes\hat s_{i_n}\}$ associated with the
prior probabilities 
$\{ \xi_{i_1}\times\cdots\times\xi_{i_n}\}$, the optimum POM minimizing the
average error probability is,  
\begin{equation}
\hat{\mit\Pi}_{i_1\cdots i_n}=\hat\pi_{i_1}\otimes\cdots\otimes\hat\pi_{i_n}
\label{eqn:threshold_pi_error}
\end{equation}
and the minimum average error probability is given as, 
\begin{equation}
P_e({\rm opt})=1-({\rm Tr}\hat\upsilon)^n, 
\label{eqn:threshold_Pe}
\end{equation}
where $\hat\upsilon\equiv\displaystyle\sum_i\xi_i\hat\pi_i$ is a Lagrange
operator.

Its proof is given in Appendix B \cite{Holevo97_private_commun}. It should
be noted that the optimum POM for the letter states $\{\hat\pi_i\}$ need
not, in general, to coincide with the one for the mutual information,
$\{\hat\pi_i^\ast\}$ in Eq. (\ref{eqn:threshold_pi_inf}), even for the same
ensemble 
$\{\hat s_i\}$. 
The result of this theorem will be discussed later with the example of the
channel coding.

The simple example of a channel coding showing the superadditivity in
capacity was already given by the authors in the case of the third
extension of the binary pure-state channel \cite{Sasaki97_SupAdd}. Here we
generalize this example into $n$-th order extension, and demonstrate the
relation 
\begin{equation}
\sup_{\hat{\bf\Pi}} 
I_n(\mbox{\boldmath $\zeta$}, \hat{\bf s}^{\otimes n},\hat{\bf\Pi})
    >n C_1(\hat{\bf s}),
\end{equation}
which ensures the strict superadditivity.  
The initial channel is made of binary letters $\hat{\bf s}=\{\vert+\rangle,
\vert-\rangle\}$ whose inner product $\kappa=\langle+\vert-\rangle$ is
assumed to be real. It is well known that the first order capacity is
achieved by the symmetric channel with the detection operators
$\{\hat\pi_1,\hat\pi_2\}$ which minimizes the average error probability
\cite{FuchsPeres96,Ban97_C1,Osaki97_C1}. These operators are given as 
$\hat\pi_i=\vert\omega_i\rangle\langle\omega_i\vert$ with 
\begin{mathletters}
\begin{eqnarray}
\vert\omega_1\rangle&=&
       \sqrt {{1-p}\over{1-\kappa^2}} \vert+\rangle
      - \sqrt {{p}\over{1-\kappa^2}} \vert-\rangle, \\
\vert\omega_2\rangle&=&
        - \sqrt {{p}\over{1-\kappa^2}} \vert+\rangle
        + \sqrt {{1-p}\over{1-\kappa^2}}\vert-\rangle, 
\end{eqnarray}
\label{eqn:w1w2}
\end{mathletters}

\noindent   
where $p=(1-\sqrt{1-\kappa^2})/2$ is the minimum average error probability.
The first order capacity is given simply as, 
\begin{equation}
C_1(\hat{\bf s})=1+(1-p)\log_2 (1-p) + p\log_2 p. 
\label{eqn:c1_binary}
\end{equation}

In the $n$-th order extension, half of all the $2^n$ sequences are used as
the codeword states and are input to the channel with equal prior
probabilities. Such $2^{n-1}$ codeword states are generated in a recursive
manner from the four codeword states $\{
\vert+++\rangle,\vert+--\rangle,\vert--+\rangle,\vert-+-\rangle  \} $ in
the third order extension, where 
$\vert+--\rangle\equiv\vert+\rangle\otimes\vert-\rangle\otimes\vert-\rangle$
, etc. 
That is, defining vectors consisting of bra-state vectors of codeword states;
\begin{equation}
{\mbox{\boldmath $\gamma$}}^{(3)}\equiv
\left(
\begin{array}{c}
\langle+++\vert\cr
\langle+--\vert\cr
\langle-+-\vert\cr
\langle--+\vert
\end{array}
\right), 
\quad 
{\mbox{\boldmath $\lambda$}}^{(3)}\equiv
\left(
\begin{array}{c}
\langle---\vert\cr
\langle-++\vert\cr
\langle+-+\vert\cr
\langle++-\vert
\end{array}
\right), 
\label{eqn:3rd_codeword_vec}
\end{equation}
they are given as, 
\begin{equation}
{\mbox{\boldmath $\gamma$}}^{(n)}
=
\left(
\begin{array}{c}
\langle S_1\vert\cr
\langle S_2\vert\cr
\vdots\cr
\langle S_{2^{n-1}}\vert
\end{array}
\right)
\equiv
\left(
\begin{array}{c}
\langle+\vert\otimes{\mbox{\boldmath $\gamma$}}^{(n-1)}\cr
\langle-\vert\otimes{\mbox{\boldmath $\lambda$}}^{(n-1)}
\end{array}
\right). 
\label{eqn:nth_codeword_vec}
\end{equation}
This codeword selection can be specified by the notation
$\left[[n,n-1,2]\right]$ according to the nomenclature of coding theory.

These codewords are decoded by the square-root measurement.  The
measurement basis vectors are defined as, 
\begin{equation}
\vert\mu_i\rangle \equiv \hat\rho^{-{1\over2}} \vert S_i\rangle, \quad
\hat\rho\equiv\sum_{i=1}^{2^{n-1}} \vert S_i\rangle\langle S_i\vert,  
\label{eqn:SqRt_basis vectors}
\end{equation}
where the prior probabilities are not included in the density matrix
$\hat\rho$ unlike Eq. (\ref{eqn:sqrt-mu}), simply for a mathematical
convenience.  As it will become  clear soon, 
these basis vectors effect the {\it optimum collective decoding} for the
above codewords.  
We have to evaluate the channel matrix $P(j\vert i)=\vert\langle\mu_j\vert
S_i\rangle\vert^2$.   
The Gram matrix of the codewords is given by, 
\begin{mathletters}
\begin{equation}
{\bf\Gamma}^{(n)}={\mbox{\boldmath $\gamma$}}^{(n)}
                      \cdot{\mbox{\boldmath $\gamma$}}^{(n)\dagger}
=\left(
\begin{array}{cc}
{\bf\Gamma}^{(n-1)}               & \kappa^2{\bf\Lambda}^{(n-1)} \cr
\kappa^2{\bf\Lambda}^{(n-1)} & {\bf\Gamma}^{(n-1)}
\end{array}
\right),
\label{eqn:Gram1}
\end{equation}
where
\begin{equation}
{\bf\Lambda}^{(n-1)}={1\over\kappa}
                               {\mbox{\boldmath $\gamma$}}^{(n-1)}
                      \cdot{\mbox{\boldmath $\lambda$}}^{(n-1)\dagger}
=\left(
\begin{array}{cc}
{\bf\Gamma}^{(n-2)}  & {\bf\Lambda}^{(n-2)} \cr
{\bf\Lambda}^{(n-2)} & {\bf\Gamma}^{(n-2)}
\end{array}
\right),
\label{eqn:Gram2}
\end{equation}
and
\begin{equation}
{\bf\Gamma}^{(3)}
=\left(
\begin{array}{cccc}
1 & \kappa^2 & \kappa^2 & \kappa^2 \cr
\kappa^2 & 1 & \kappa^2 & \kappa^2 \cr
\kappa^2 & \kappa^2 & 1 & \kappa^2 \cr
\kappa^2 & \kappa^2 & \kappa^2 & 1
\end{array}
\right),
\label{eqn:Gram4}
\end{equation}
\begin{equation}
{\bf\Lambda}^{(3)}=\left(
\begin{array}{cccc}
\kappa^2 & 1  & 1  &  1   \cr
 1  & \kappa^2 & 1  &  1   \cr
 1  & 1  & \kappa^2  &  1   \cr
 1  & 1  & 1  & \kappa^2
\end{array}
\right). 
\label{eqn:Gram5}
\end{equation}
\end{mathletters}

\noindent 
${\bf\Gamma}^{(n)}$ and ${\bf\Lambda}^{(n)}$ can be diagonalized by a
$2^{n-1}\times2^{n-1}$ matrix
\begin{mathletters}
\begin{equation}
{\bf Q}^{(n)}={\bf H}_{2^{n-1}}/\sqrt{2^{n-1}}
\end{equation}
where ${\bf H}_{2^{n-1}}$ is the Hadamard matrix defined by 
\begin{equation}
{\bf H}_{2k}
=\left(
\begin{array}{cc}
{\bf H}_{k}  & {\bf H}_{k} \cr
{\bf H}_{k} & -{\bf H}_{k}
\end{array}
\right), \quad
{\bf H}_{2}
=\left(
\begin{array}{cc}
1  & 1 \cr
1 & -1
\end{array}
\right). 
\end{equation}
\end{mathletters}

\noindent    
The diagonalized matrices, 
\begin{mathletters}
\begin{eqnarray}
{\bf G}^{(n)}\equiv{\bf Q}^{(n)\dagger}{\bf\Gamma}^{(n)}{\bf Q}^{(n)}, \\
{\bf F}^{(n)}\equiv{\bf Q}^{(n)\dagger}{\bf\Lambda}^{(n)}{\bf Q}^{(n)}, 
\end{eqnarray}
\end{mathletters}

\noindent    
can be decomposed into $2^{n-2}\times2^{n-2}$ matrices ${\bf G}^{(n-1)}$
and ${\bf F}^{(n-1)}$ as, 
\begin{mathletters}
\begin{equation}
{\bf G}^{(n)}=
\left(
\begin{array}{cc}
{\bf G}^{(n-1)}+\kappa^2{\bf F}^{(n-1)}  & 0 \cr
0 & {\bf G}^{(n-1)}-\kappa^2{\bf F}^{(n-1)}
\end{array}
\right), 
\end{equation}
\begin{equation}
{\bf F}^{(n)}=
\left(
\begin{array}{cc}
{\bf G}^{(n-1)}+{\bf F}^{(n-1)}  & 0 \cr
0 & {\bf G}^{(n-1)}-{\bf F}^{(n-1)}
\end{array}
\right). 
\end{equation}
\end{mathletters}

\noindent 
After recursive decompositions, they can be represented as, 
\begin{mathletters}
\begin{equation}
{\bf G}^{(n)}=
\left(
\begin{array}{ccc}
{\bf A}(n,1)  & {} & {} \cr
{} & \ddots & {} \cr
{} & {} & {\bf A}(n,2^{n-3})
\end{array}
\right), 
\end{equation}
\begin{equation}
{\bf F}^{(n)}=
\left(
\begin{array}{ccc}
{\bf B}(n,1)  & {} & {} \cr
{} & \ddots & {} \cr
{} & {} & {\bf B}(n,2^{n-3})
\end{array}
\right), 
\end{equation}
\end{mathletters}

\noindent  
where ${\bf A}(n,k)$ and ${\bf B}(n,k)$ $(k=1,\cdots,2^{n-3})$ are
$4\times4$ block matrices defined by, 
\begin{mathletters}
\begin{eqnarray}
{\bf A}(n,k)=a(n,k){\bf G}^{(3)} + b(n,k){\bf F}^{(3)}, \\
{\bf B}(n,k)=c(n,k){\bf G}^{(3)} + d(n,k){\bf F}^{(3)},
\end{eqnarray}
\label{eqn:AB_GF}
\end{mathletters}

\noindent
with 
\begin{mathletters}
\begin{equation}
{\bf G}^{(3)}=
\left(
\begin{array}{cccc}
1+3\kappa^2  & {} & {} & {} \cr
{} & 1-\kappa^2 & {} & {} \cr
{} & {} & 1-\kappa^2 & {} \cr
{} & {} & {} & 1-\kappa^2
\end{array}
\right), 
\end{equation}
\begin{equation}
{\bf F}^{(3)}=
\left(
\begin{array}{cccc}
3+\kappa^2  & {} & {} & {} \cr
{} & -1+\kappa^2 & {} & {} \cr
{} & {} & -1+\kappa^2 & {} \cr
{} & {} & {} & -1+\kappa^2
\end{array}
\right).  
\end{equation}
\end{mathletters}

\noindent 
The coefficients in Eq. (\ref{eqn:AB_GF}) are determined by the following
recursive formula for $k=1,\cdots,2^{n-4}$, 
\begin{mathletters}
\begin{equation}
\begin{array}{lll}
a(n,k)             &=&a(n-1,k) + \kappa^2 c(n-1,k), \\
a(n,2^{n-4}+k)&=&a(n-1,k) - \kappa^2 c(n-1,k), \\
b(n,k)             &=&b(n-1,k) + \kappa^2 d(n-1,k), \\
b(n,2^{n-4}+k)&=&b(n-1,k) - \kappa^2 d(n-1,k), \\
c(n,k)             &=&a(n-1,k) +   c(n-1,k), \\
c(n,2^{n-4}+k)&=&a(n-1,k) -   c(n-1,k), \\
d(n,k)             &=&b(n-1,k) +   d(n-1,k), \\
d(n,2^{n-4}+k)&=&b(n-1,k) -   d(n-1,k), 
\end{array}
\end{equation}
with the initial values,  
\begin{equation}
\begin{array}{lll}
a(4,1)&=&a(4,2)=1, \\
b(4,1)&=&-b(4,2)=\kappa^2, \\
c(4,1)&=&c(4,2)=d(4,1)=-d(4,2)=1. 
\end{array}
\end{equation}
\end{mathletters}

\noindent 
Thus the diagonal matrices ${\bf A}(n,k)$'s are obtained, and the
square-root of the Gram matrix is given as, 
\begin{equation}
{  \sqrt{  {\bf\Gamma}^{(n)}  }  }
=
{\bf Q}^{(n)} 
\left(
\begin{array}{ccc}
\sqrt{{\bf A}(n,1)}  & {} & {} \cr
{} & \ddots & {} \cr
{} & {} & \sqrt{{\bf A}(n,2^{n-3})}
\end{array}
\right)  {\bf Q}^{(n)\dagger}. 
\end{equation}
For representing the result, let us define, 
\begin{mathletters}
\begin{eqnarray}
\alpha(n,k)&\equiv&\sqrt{  (1+3\kappa^2) a(n,k) + (3+\kappa^2) b(n,k) }, \\
\beta  (n,k)&\equiv&\sqrt{  (1-\kappa^2) [a(n,k) - b(n,k) ]  },  
\end{eqnarray}
and
\begin{eqnarray}
{\bf D}(n,k)&\equiv&{\bf Q}^{(3)}\sqrt{{\bf A}(n,k)}{\bf Q}^{(3)\dagger}\\
&=&
\left(
\begin{array}{cccc}
\mu(n,k)  & \nu(n,k) & \nu(n,k) & \nu(n,k) \cr
\nu(n,k) & \mu(n,k) & \nu(n,k) & \nu(n,k) \cr
\nu(n,k) & \nu(n,k) & \mu(n,k) & \nu(n,k) \cr
\nu(n,k) & \nu(n,k) & \nu(n,k) & \mu(n,k)
\end{array}
\right), 
\label{eqn:D(n,k)}
\end{eqnarray}
where 
\begin{eqnarray}
\mu(n,k)&\equiv&{1\over4}[\alpha(n,k) + 3\beta(n,k)], \\
\nu(n,k)&\equiv&{1\over4}[\alpha(n,k) -   \beta(n,k)].   
\end{eqnarray}
\end{mathletters}

\noindent 
Then ${  \sqrt{  {\bf\Gamma}^{(n)}  }  }$ can be represented as, 
\begin{mathletters}
\begin{equation}
{  \sqrt{  {\bf\Gamma}^{(n)}  }  }
=
\left(
\begin{array}{lllll}
{\bf R}(n,1) & {\bf R}(n,2) & {\bf R}(n,3) & {\bf R}(n,4) & \cdots \cr
{\bf R}(n,2) & {\bf R}(n,1) & {\bf R}(n,4) & {\bf R}(n,3) & \cdots \cr
{\bf R}(n,3) & {\bf R}(n,4) & {\bf R}(n,1) & {\bf R}(n,2) & \cdots \cr
{\bf R}(n,4) & {\bf R}(n,3) & {\bf R}(n,2) & {\bf R}(n,1) & \cdots \cr
\vdots         & \vdots        & \vdots         & \vdots        & \vdots \cr    
{\bf R}(n,2^{n-3}-1)&{\bf R}(n,2^{n-3})&{\bf R}(n,2^{n-3}-3)&{\bf
R}(n,2^{n-3}-2)& \cdots \cr   
{\bf R}(n,2^{n-3})&{\bf R}(n,2^{n-3}-1)&{\bf R}(n,2^{n-3}-2)&{\bf
R}(n,2^{n-3}-3)& \cdots   
\end{array}
\right),
\end{equation}
where 
\begin{equation}
\left(
\begin{array}{l}
{\bf R}(n,1)\cr
\vdots\cr
{\bf R}(n,2^{n-3})
\end{array}
\right)
\equiv
{1\over{2^{n-3}}}  {\bf H}_{2^{n-1}}
\left(
\begin{array}{l}
{\bf D}(n,1)\cr
\vdots\cr
{\bf D}(n,2^{n-3})
\end{array}
\right). 
\end{equation}
\label{eqn:SqRt_GramMat}
\end{mathletters}

\noindent 
${\bf R}(n,k)$ can be further arranged in the following form, 
\begin{equation}
{\bf R}(n,k)=
\left(
\begin{array}{cccc}
u(n,k)  & v(n,k) & v(n,k) & v(n,k) \cr
v(n,k) & u(n,k) & v(n,k) & v(n,k) \cr
v(n,k) & v(n,k) & u(n,k) & v(n,k) \cr
v(n,k) & v(n,k) & v(n,k) & u(n,k)
\end{array}
\right). 
\label{eqn:R(n,k)_uv}
\end{equation}
The two kinds of components $u(n,k)$ and $v(n,k)$ can be calculated by 
\begin{mathletters}
\begin{equation}
\left(
\begin{array}{l}
u(n,1)\cr
\vdots\cr
u(n,2^{n-3})
\end{array}
\right)
\equiv
{1\over{2^{n-3}}}  {\bf H}_{2^{n-3}}
\left(
\begin{array}{l}
\mu(n,1)\cr
\vdots\cr
\mu(n,2^{n-3})
\end{array}
\right), 
\label{eqn:u}
\end{equation}
\begin{equation}
\left(
\begin{array}{l}
v(n,1)\cr
\vdots\cr
v(n,2^{n-3})
\end{array}
\right)
\equiv
{1\over{2^{n-3}}}  {\bf H}_{2^{n-3}}
\left(
\begin{array}{l}
\nu(n,1)\cr
\vdots\cr
\nu(n,2^{n-3})
\end{array}
\right). 
\label{eqn:v}
\end{equation}
\end{mathletters}

\noindent 
After squaring each component of $\sqrt{  {\bf\Gamma}^{(n)}  }$, the
channel matrix $P(j \vert i)$ can be obtained. According to the symmetry
seen in Eq. (\ref{eqn:SqRt_GramMat}), it is easy to see that the mutual
information is given as, 
\begin{equation}
I_n(S:\mu)=n-1 + \sum_{k=1}^{2^{n-3}} 
\left[   u(n,k)^2\log u(n,k)^2 + 3 v(n,k)^2\log v(n,k)^2   \right].  
\label{eqn:I_n}
\end{equation}

In order to see the quantum gain, the difference between the mutual
information per letter state 
$I_n(S:\mu)/n$ and $C_1(\hat{\bf s})$ is plotted as functions of $\kappa$ 
in Fig. \ref{fig2}.  The dashed line corresponds to the case of $n=2$ where
the two codeword states  
$\{ \vert++\rangle ,  \vert--\rangle \}$ are sent with the same prior
probabilities and are detected by the optimum measurement minimizing the
average error probability. In this case, the positive quantum gain was not
found in the whole region of $\kappa$.  
For $n=3\sim13$ (solid lines), the difference becomes positive at the
larger side of $\kappa$. 
This positive gain clearly shows the superadditivity in capacity. 
Let $\kappa_\ast$ be the value of $\kappa(<1)$ for which the difference
becomes zero. 
Then for $\kappa_\ast<\kappa<1$ the difference is always positive, and 
as $n$ increases, $\kappa_\ast$ decreases so that the positive gain appears
in wider region of $\kappa$. 
This relation is plotted in Fig. \ref{fig3}. The circles represent the
points $(\kappa_\ast,n)$. The solid line is just a guide for eye.  The
dashed line corresponds to the curve of $n=2\kappa^{-1.5}$. This figure may
provide a rough estimate for $n$ in order to obtain the positive gain. That
is,  for a given $\kappa$, one may guess that  the superadditivity will
appear when the order of extension for our $\left[[n,n-1,2]\right]$ code is
taken as an integer $n$ larger than $2\kappa^{-1.5}$. Unfortunately we did
not  succeed in giving more rigorous condition. 
The maximum amount of the positive gain is still quantitatively unsatisfactory  
compared with the gap between the intrinsic capacity $C(\hat{\bf s})$ and
the first order capacity $C_1(\hat{\bf s})$.  Actually it is less than 10\%
of the maximum gap. In the case of $n=9$ for which the maximum gain was
obtained, 
$C(\hat{\bf s})$, $I_9(S:\mu)/9$ and $C_1(\hat{\bf s})$ versus $\kappa$ are
plotted in Fig. \ref{fig4}.

As far as the minimum average error probability is concerned, the
square-root measurement we used (Eq. (\ref{eqn:SqRt_basis vectors})) is the
optimum for our $\left[[n,n-1,2]\right]$ code  (Eq.
(\ref{eqn:nth_codeword_vec}))  because all of the diagonal components of
$\sqrt{  {\bf\Gamma}^{(n)}  }$ are equal to  $u(n,1)$ as seen from Eqs. 
(\ref{eqn:SqRt_GramMat}) and (\ref{eqn:R(n,k)_uv}),  for  which Theorem 1
holds. 
The minimum average error probabilities versus $\kappa$ are plotted for
$n=3$, 5, 7, 9, 11, 13 by the solid lines in Fig. \ref{fig5}. 
For a fixed $\kappa$, they increase as with $n$.  
The dotted lines represent the minimum average error probabilities
corresponding to the {\it threshold points}, that is,  Eq.
(\ref{eqn:threshold_Pe}).  
Although the error probabilities of our code $\left[[n,n-1,2]\right]$ are 
smaller  than those of the {\it threshold points}, they are  still larger
than $p=(1-\sqrt{1-\kappa^2})/2$ (the minimum error of the initial channel)
at larger side of $\kappa$.   In spite of this, the quantum gain 
$I_n(S:\mu)/n > C_1(\hat{\bf s})$ reveals itself within such regions.

The same tendency could be seen in other codes. 
Let us consider  the so-called simplex code
$\left[[2^r-1,r,2^{r-1}]\right]$, for example.  All the codewords are the
same distance apart.  Let $n=2^r-1$ and $M=2^r$. Suppose that $M$-ary
codewords are used with equal prior probabilities. Then the square-root
measurement is again the optimum collective decoding for them. 
Defining,
\begin{mathletters}
\begin{eqnarray}
\alpha(n)&\equiv&\sqrt{  1 + (M-1) \kappa^{M/2} }, \\
\beta(n)  &\equiv&\sqrt{  1 - \kappa^{M/2} },  
\end{eqnarray}
\end{mathletters}

\noindent 
it is straightforward to see, 
\begin{mathletters}
\begin{eqnarray}
(\sqrt{  {\bf\Gamma}^{(n)}  })_{ii}&=& 
u(n)\equiv{1\over M}\left[\alpha(n)+(m-1)\beta(n)\right], \\
(\sqrt{  {\bf\Gamma}^{(n)}  })_{ij}&=&  
v(n)\equiv{1\over M}\left[\alpha(n)-\beta(n)\right], \quad i\ne j.  
\end{eqnarray}
\end{mathletters}
The mutual information is then given by, 
\begin{equation}
I_n(S:\mu)=\log M + u(n)^2\log u(n)^2 + (M-1) v(n)^2\log v(n)^2.  
\label{eqn:simplex_I_n}
\end{equation}
This code is compared with the previous one  at $n=7$ in terms of both the
mutual information per letter states and the minimum average error
probability in Fig. \ref{fig6} and \ref{fig7}, respectively.   The
$\left[[7,3,4]\right]$ simplex code has higher distinguishability of the
codewords  than the  $\left[[7,6,2]\right]$  code so that the minimum
average error is much smaller, while its mutual information is not
necessarily larger than the latter. The former overcomes the latter  only
in the region $0.82<\kappa<1$.  Around this region, the minimum average
error probability is, again,  larger than the one in the initial channel
$p$.

This tendency may be understood as  a result from the facts that in order
to produce the quantum gain a quantum interference among the codeword
states must occur to reduce certain components of the channel matrix, and
that such  a quantum interference occurs more drastically when the
nonorthogonality of the codeword states is larger, hence in larger side of
$\kappa$.  On the other hand,  the nonorthogonality causes the certain
amount of decoding error as well.  This decreases the amount of
transmittable information, while the quantum gain may appear if the quantum
interference reduces  certain components of the channel matrix in a proper
manner.     
Thus at shorter block length, the quantum gain as the difference
$I_n-C_1>0$ is likely to appear in larger side of $\kappa$ being
accompanied by a certain amount of decoding error.  A rough guide in order
to construct a channel attaining the quantum gain for a given $\kappa$, is
as follows; the ratio of the number of the message bits $k$ to the block
length $n$ should be taken larger than  $C_1(\hat{\bf s})$ at this $\kappa$
first, and then  codeword states should be selected being with distance as
equally apart as possible. 
As $\kappa$ becomes closer to the unity, it becomes more effective in
obtaining the quantum gain to take the ratio $k/n$ small and to select the
codeword states  being   distant. As an example, the simplex code
$[[7,3,4]]$ is compared with the code $[[3,2,2]]$ in the region
$0.75<\kappa<1$ in Fig. \ref{fig8}. Around at $\kappa=0.7$, the $[[7,3,4]]$
code (solid line) is more efficient in terms of the mutual information than
the  $[[3,2,2]]$ code (dotted line). 
For constructing codes such that the decoding error can be as small as
possible and the rate can reach the Holevo bound, a larger block length at
which the typical subspace can be well defined is necessary. 
Practical methods for obtaining larger quantum gain must be studied in
great detail along this direction.

\section{Concluding remarks}\label{sec5}

The initial channel considered in this paper is the binary symmetric
pure-state channel which is the simplest quantum channel. When the binary
letter states are orthogonal, the channel is error free, and there is no
quantum regime, that is, $n$-th order capacity obviously satisfies the
strict additivity $C_n=nC_1$.  When they are nonorthogonal, i.e.
$0<\kappa<1$, the strict {\it superadditivity} $C_n>nC_1$, in turn, reveals
itself. What we have shown in this paper is a demonstration of $I_n>nC_1$
that ensures the strict {\it superadditivity}.

The first part of this paper was devoted to the optimum collective decoding
of the codeword states at the minimum average error probability. The scheme
we proposed consists of  the unitary transformation and the separate
measurement.  The unitary transformation generates appropriate
superposition states among the codeword states such that the minimum
average error is attained at the separate measurement.  These output states
are not separable into the letter states any more. Minimization of decoding
error is just a manifestation of the optimum use of quantum interference
associated with this kind of superposition states. The required unitary
transformation is essentially a conditional dynamics in a higher
dimensional Hilbert space, and is realized by a quantum circuit which is
capable of manipulating each letter state in a conditional manner depending
on the other letter states.  Thus our scheme suggests a state-of-the-art
quantum decoder structure.

Quantum channels involving the above collective decoding has a {\it memory
effect}, i.e.  
$P( y_1 y_2 \cdots y_n \vert x_1 x_2 \cdots x_n) 
\ne \displaystyle\prod_{i=1}^n P( y_i \vert x_i)$.   
This inseparability of quantum channel is a direct origin of the quantum
gain $I_n>nC_1$. 
Only when codeword states are selected suitably, this inseparability leads
to the quantum gain.  We gave a heuristic approach  to attain this gain for
a given letter-state ensemble.  Our examples are always accompanied by  a
larger amount of decoding error than the minimum average error in the
initial channel. As mention in the previous section, this dilemma is
because both decoding error and the quantum gain are originated from the
nonorthogonality of the letter states.

Although some basic aspects for realizing the quantum gain were clarified
by this paper,  practical codes that transmit classical alphabet faithfully
at the maximum rate are still completely unknown. Even in classical
information theory, realization of such codes that achieve asymptotically
error free transmission at the rate $C_1$ is very difficult. It might be an
interesting problem  to consider an application of some  conventional error
correcting codes to the $n$-product quantum channels described in this
paper. This will lead to realization of asymptotically error free
transmission at the rate $I_n/n$ ($>C_1$).  For the ultimate quantum
channel coding, typicality of the Hilbert space spanned by the codeword
states and sophisticated quantum error-correction might be considered
together.

\acknowledgements

The authors would like to thank  Dr. Holevo of Steklov Mathematical
Institute and Dr. Usuda of Nagoya institute of technology  for giving
crucial comments on this work.  They would also like to thank  Dr. C. A.
Fuchs of California Institute of technology, Dr. C. H. Bennett of IBM T. J.
Watson research center, Dr. M. Ban of Hitachi Advanced Research Laboratory,
Drs. K. Yamazaki and M. Osaki of Tamagawa University, for their helpful
discussions.

\appendix
\section{Proof of the theorem 2}

We first prove the following lemma. 

\noindent
{\bf Lemma}

\begin{eqnarray}
\sup_{\hat{\bf\Pi}} 
&{}&I(\mbox{\boldmath $\xi$}^{(1)}\otimes\mbox{\boldmath $\xi$}^{(2)},
       \hat{\bf s}^{(1)}\otimes\hat{\bf s}^{(2)}:\hat{\bf\Pi})
\nonumber\\
&=&
\sup_{ \hat{\mbox{\boldmath $\pi$}}^{(1)} } 
I(\mbox{\boldmath $\xi$}^{(1)},
       \hat{\bf s}^{(1)}:\hat{\mbox{\boldmath $\pi$}}^{(1)})
+\sup_{ \hat{\mbox{\boldmath $\pi$}}^{(2)} } 
I(\mbox{\boldmath $\xi$}^{(2)},
       \hat{\bf s}^{(2)}:\hat{\mbox{\boldmath $\pi$}}^{(2)}). 
\end{eqnarray}

\noindent
{\bf Proof of lemma}

1) Clearly, 

\begin{eqnarray}
\sup_{\hat{\bf\Pi}} 
&{}&I(\mbox{\boldmath $\xi$}^{(1)}\otimes\mbox{\boldmath $\xi$}^{(2)},
       \hat{\bf s}^{(1)}\otimes\hat{\bf s}^{(2)}:\hat{\bf\Pi})
\nonumber\\
&\ge&
\sup_{ \hat{\bf\Pi}=\hat{\mbox{\boldmath
$\pi$}}^{(1)}\otimes\hat{\mbox{\boldmath $\pi$}}^{(2)} } 
I(\mbox{\boldmath $\xi$}^{(1)}\otimes\mbox{\boldmath $\xi$}^{(2)},
       \hat{\bf s}^{(1)}\otimes\hat{\bf s}^{(2)}:\hat{\bf\Pi}) 
\nonumber \\
&=&
\sup_{ \hat{\mbox{\boldmath $\pi$}}^{(1)} } 
I(\mbox{\boldmath $\xi$}^{(1)},
       \hat{\bf s}^{(1)}:\hat{\mbox{\boldmath $\pi$}}^{(1)})
+
\sup_{ \hat{\mbox{\boldmath $\pi$}}^{(2)} } 
I(\mbox{\boldmath $\xi$}^{(2)},
       \hat{\bf s}^{(2)}:\hat{\mbox{\boldmath $\pi$}}^{(2)}), 
\label{eqn:L-1}
\end{eqnarray}

2) Representing 
$P(j\vert i_1, i_2)={\rm Tr}(\hat\Pi_j\hat s_{i_1}^{(1)}\otimes\hat
s_{i_2}^{(2)})$, 
where $\hat\Pi_j$ is a POM on ${\cal H}_s^{(1)}\otimes{\cal H}_s^{(2)}$,  
\begin{eqnarray}
&{}&I(\mbox{\boldmath $\xi$}^{(1)}\otimes\mbox{\boldmath $\xi$}^{(2)},
       \hat{\bf s}^{(1)}\otimes\hat{\bf s}^{(2)}:\hat{\bf\Pi})
\nonumber\\
&=&
\sum_j\sum_{i_1,i_2}  \xi_{i_1}^{(1)} \xi_{i_2}^{(2)}  P(j\vert i_1, i_2) 
\log\Big[{
               {P(j\vert i_1, i_2)}
               \over
               {\sum_{k_1,k_2}\xi_{k_1}^{(1)} \xi_{k_2}^{(2)}P(j\vert k_1,
k_2)} 
               }\Big] 
\nonumber \\
&=&
\sum_j\sum_{i_1,i_2}  \xi_{i_1}^{(1)} \xi_{i_2}^{(2)}  P(j\vert i_1, i_2) 
\nonumber\\
&\Big\{&
\log\Big[{
               {P(j\vert i_1, i_2)}
               \over
               {\sum_{k_1}\xi_{k_1}^{(1)} P(j\vert k_1, i_2)} 
               }\Big]
\nonumber\\  
&+&
\log\Big[{
               {\sum_{k_1}\xi_{k_1}^{(1)} P(j\vert k_1, i_2)}
               \over
               {\sum_{k_1,k_2}\xi_{k_1}^{(1)} \xi_{k_2}^{(2)}P(j\vert k_1,
k_2)} 
               }\Big]  
\Big\}
\nonumber \\
&=&
\sum_{i_2}\xi_{i_2}^{(2)}
\Big\{
\sum_j\sum_{i_1}  \xi_{i_1}^{(1)}   P(j\vert i_1, i_2) 
\log\Big[{
               {P(j\vert i_1, i_2)}
               \over
               {\sum_{k_1}\xi_{k_1}^{(1)} P(j\vert k_1, i_2)} 
               }\Big]
\Big\}
\nonumber \\
&+&
\sum_j\sum_{i_2}\xi_{i_2}^{(2)}
\left[  \sum_{i_1}  \xi_{i_1}^{(1)}   P(j\vert i_1, i_2)  \right]
\log\Big\{
               {
               \left[  \sum_{k_1}  \xi_{k_1}^{(1)}   P(j\vert k_1, i_2)  \right]
               \over
               {\sum_{k_2}\xi_{k_2}^{(2)}\left[  \sum_{k_1} 
\xi_{k_1}^{(1)}   P(j\vert k_1, i_2)  \right]} 
               }
               \Big\}.   
\label{eqn:L-2}
\end{eqnarray}
We introduce two kind of POM on ${\cal H}_s^{(1)}$ and ${\cal H}_s^{(2)}$ as, 
\begin{eqnarray}
\hat\Pi_{(i_2)j}^{(1)}
&\equiv&{\rm Tr}^{(2)} \left( \hat\Pi_j \hat I^{(1)}\otimes\hat
s_{i_2}^{(2)} \right), \\
\bar\Pi_{j}^{(2)}
&\equiv&\sum_{i_1}\xi_{i_1}^{(1)}
{\rm Tr}^{(1)} \left( \hat\Pi_j \hat s_{i_1}^{(1)} \otimes\hat I^{(2)} \right), 
\end{eqnarray}
where ${\rm Tr}^{(i)}$ means taking the trace over the space ${\cal
H}_s^{(i)}$, and $\hat I^{(i)}$ is the identity operators in ${\cal
H}_s^{(i)}$. 
Representing the conditional probabilities in Eq. (\ref{eqn:L-2}) as, 
\begin{eqnarray}
P(j\vert i_1, i_2)
&=&{\rm Tr}^{(1)} \left(  \hat\Pi_{(i_2)j}^{(1)}   \hat s_{i_1}^{(1)}
\right), \\
\sum_{i_1} \xi_{i_1}^{(1)} P(j\vert i_1, i_2)
&\equiv&
{\rm Tr}^{(2)} \left( \bar\Pi_{j}^{(2)} \hat s_{i_2}^{(2)} \right) ,
\end{eqnarray}
we see that Eq. (\ref{eqn:L-2}) is equivalent to the following, 
\begin{eqnarray}
&{}&I(\mbox{\boldmath $\xi$}^{(1)}\otimes\mbox{\boldmath $\xi$}^{(2)},
       \hat{\bf s}^{(1)}\otimes\hat{\bf s}^{(2)}:\hat{\bf\Pi}) 
\nonumber\\
&=&
\sum_{ i_2} \xi_{i_2}^{(2)}
I(\mbox{\boldmath $\xi$}^{(1)},
       \hat{\bf s}^{(1)}:\hat{\bf\Pi}_{(i_2)}^{(1)})
+I(\mbox{\boldmath $\xi$}^{(2)},
       \hat{\bf s}^{(2)}:\bar{\bf\Pi}_{(2)}). 
\end{eqnarray}
Hence 
\begin{eqnarray}
&{}&\sup_{\hat{\bf\Pi}} 
I(\mbox{\boldmath $\xi$}^{(1)}\otimes\mbox{\boldmath $\xi$}^{(2)},
       \hat{\bf s}^{(1)}\otimes\hat{\bf s}^{(2)}:\hat{\bf\Pi})
\nonumber\\
&=&
\sum_j \xi_j^{(2)}
\sup_{\hat{\bf\Pi}}
I(\mbox{\boldmath $\xi$}^{(1)},
       \hat{\bf s}^{(1)}:\hat{\bf\Pi}_j^{(1)})
+
\sup_{\hat{\bf\Pi}}
I(\mbox{\boldmath $\xi$}^{(2)},
       \hat{\bf s}^{(2)}:\bar{\bf\Pi}_{(2)}) 
\nonumber\\
&\le&
\sup_{\hat{\mbox{\boldmath $\pi$}}^{(1)}}
I(\mbox{\boldmath $\xi$}^{(1)},
       \hat{\bf s}^{(1)}:\hat{\mbox{\boldmath $\pi$}}^{(1)})
+
\sup_{\hat{\mbox{\boldmath $\pi$}}^{(2)}}
I(\mbox{\boldmath $\xi$}^{(2)},
       \hat{\bf s}^{(2)}:\hat{\mbox{\boldmath $\pi$}}^{(2)}). 
\label{eqn:L-3}
\end{eqnarray}
The two inequalities (\ref{eqn:L-1}) and (\ref{eqn:L-3}) prove the lemma. 

Now suppose that 
$I(\mbox{\boldmath $\xi$}^{(i)},\hat{\bf s}^{(i)}:\hat{\mbox{\boldmath
$\pi$}}^{(i)})$ is maximized when $\mbox{\boldmath
$\xi$}^{(i)}=\mbox{\boldmath $\xi$}_\ast^{(i)}$ and 
$\hat{\mbox{\boldmath $\pi$}}^{(i)}=\hat{\mbox{\boldmath
$\pi$}}_\ast^{(i)}$. Then the lemma means that 
\begin{eqnarray}
&{}&\sup_{\hat{\bf\Pi}} 
I(\mbox{\boldmath $\xi$}_\ast^{(1)}\otimes\mbox{\boldmath $\xi$}_\ast^{(2)},
       \hat{\bf s}^{(1)}\otimes\hat{\bf s}^{(2)}:\hat{\bf\Pi})
\nonumber\\
&=&C_1(\hat{\bf s}^{(1)}) + C_1(\hat{\bf s}^{(2)}). 
\label{eqn:L-4}
\end{eqnarray}
On the other hand, 
\begin{eqnarray}
&{}& 
I(\mbox{\boldmath $\xi$}_\ast^{(1)}\otimes\mbox{\boldmath $\xi$}_\ast^{(2)},
       \hat{\bf s}^{(1)}\otimes\hat{\bf s}^{(2)}
       :\hat{\mbox{\boldmath $\pi$}}_\ast^{(1)}\otimes\hat{\mbox{\boldmath
$\pi$}}_\ast^{(2)})
\nonumber\\
&=&I(\mbox{\boldmath $\xi$}_\ast^{(1)},\hat{\bf
s}^{(1)}:\hat{\mbox{\boldmath $\pi$}}_\ast^{(1)}) 
+ I(\mbox{\boldmath $\xi$}_\ast^{(2)},\hat{\bf
s}^{(2)}:\hat{\mbox{\boldmath $\pi$}}_\ast^{(2)}) 
\nonumber\\
&=&C_1(\hat{\bf s}^{(1)}) + C_1(\hat{\bf s}^{(2)}). 
\label{eqn:L-5}
\end{eqnarray}
These two equations prove the theorem. \hfill$\Box$

\section{Proof of the theorem 3}

The necessary and sufficient condition that  $\{\hat\pi_i\}$ is the optimum
POM are 
\begin{itemize}
\item[i)] $\hat\pi_j(\hat w_j - \hat w_k)\hat\pi_k =0, \quad \forall (j,k),$
\item[ii)] $\hat\upsilon-\hat w_j \ge0, \quad \forall j,$
\end{itemize}
where $\hat w_i=\xi_i\hat s_i$ and $\hat\upsilon=\displaystyle\sum_j \hat
w_j\hat\pi_j$ are the risk operators and the Lagrange operator,
respectively.
We would like to prove that $\hat\Pi_{j_1\cdots j_l}$
$(=\hat\pi_{j_1}\otimes\cdots\otimes\hat\pi_{j_l})$ satisfy 
\begin{itemize}
\item[i)$'$] $\hat\Pi_{j_1\cdots j_l}(\hat W_{j_1\cdots j_l} - \hat
W_{k_1\cdots k_l})\hat\Pi_{k_1\cdots k_l} =0, $
\item[ii)$'$] $\hat\Upsilon-\hat W_{j_1\cdots j_l} \ge0, $
\end{itemize}
where $\hat W_{j_1\cdots j_l} =\hat w_{j_1}  \otimes\cdots\otimes \hat
w_{j_l}$ and 
$\hat\Upsilon=\displaystyle\sum_{j_1\cdots j_l} \hat W_{j_1\cdots
j_l}\hat\Pi_{j_1\cdots j_l} = \hat\upsilon^{\otimes n}$.  
Here note that the following formulas: 
\begin{equation}
\begin{array}{lll}
A_1&\otimes&\cdots\otimes A_l - B_1\otimes\cdots\otimes B_l \\
{}&=&(A_1-B_1)\otimes A_2\otimes\cdots\otimes A_l \\
{}&+&B_1\otimes(A_2-B_2) \otimes\cdots\otimes A_l + \cdots\\
{}&+&B_1\otimes B_2\otimes\cdots\otimes (B_l-A_l).  
\end{array}
\label{AB}
\end{equation} 
Then to ensure i)$'$, we rearrange the left-hand side as 
\begin{equation}
\begin{array}{lll}
&{}&\hat\Pi_{j_1\cdots j_l}(\hat W_{j_1\cdots j_l} - \hat W_{k_1\cdots
k_l})\hat\Pi_{k_1\cdots k_l} \\
{}&=&\hat\pi_{j_1}\hat w_{j_1} \hat\pi_{k_1} \otimes\cdots\otimes
\hat\pi_{j_n}\hat w_{j_n} \hat\pi_{k_n} \\
{}&-&\hat\pi_{j_1}\hat w_{k_1} \hat\pi_{k_1} \otimes\cdots\otimes
\hat\pi_{j_n}\hat w_{k_n} \hat\pi_{k_n}, 
\end{array}
\label{PW}
\end{equation} 
and put $A_m=\hat\pi_{j_m}\hat w_{j_m} \hat\pi_{k_m}$  and 
$B_m=\hat\pi_{j_m}\hat w_{k_m} \hat\pi_{k_m}$. 
Since i) is equivalent to $A_m-B_m=0$, after Eq. (\ref{AB}) is applied to
Eq. (\ref{PW}) we obtain i)$'$. Similarly, to show ii)$'$, we put
$A_m=\hat\upsilon$ and  $B_m=\hat w_{j_m}$. From the definitions, $A_m\ge0$
and $B_m\ge0$. In addition, ii) is nothing but $A_m\ge B_m$. So when
$\hat\Upsilon-\hat W_{j_1\cdots j_l} $ is decomposed by Eq. (\ref{AB}), its
nonnegative definiteness is obvious.  
\hfill$\Box$

\begin{figure}
\centerline{\epsffile{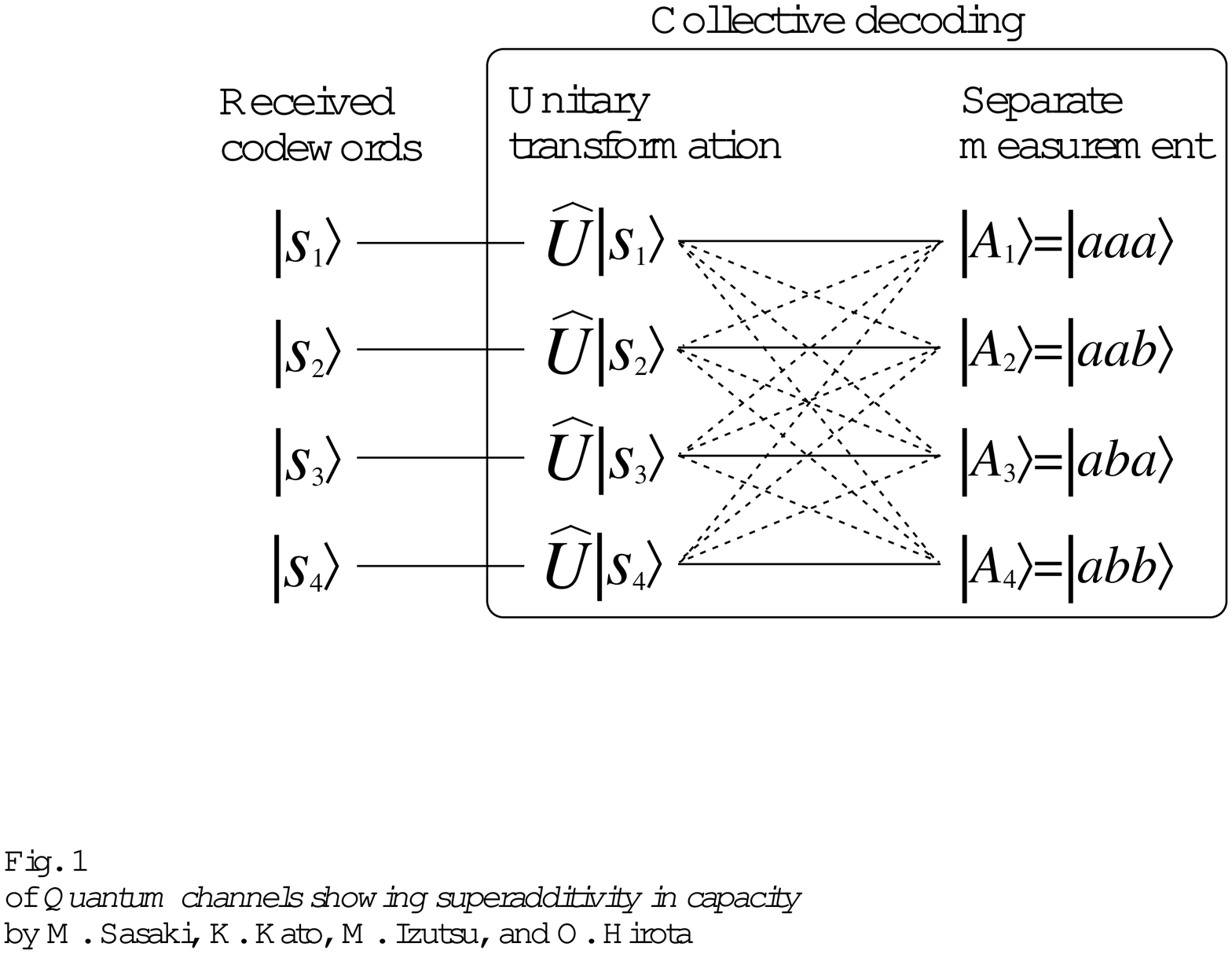}}
\caption{ The channel model obtained by decomposing the collective decoding
into the unitary transformation and the separate measurement in the case of
$n=3$ and $M=4$.   }
\label{fig1}
\end{figure}

\begin{figure}
\centerline{\epsffile{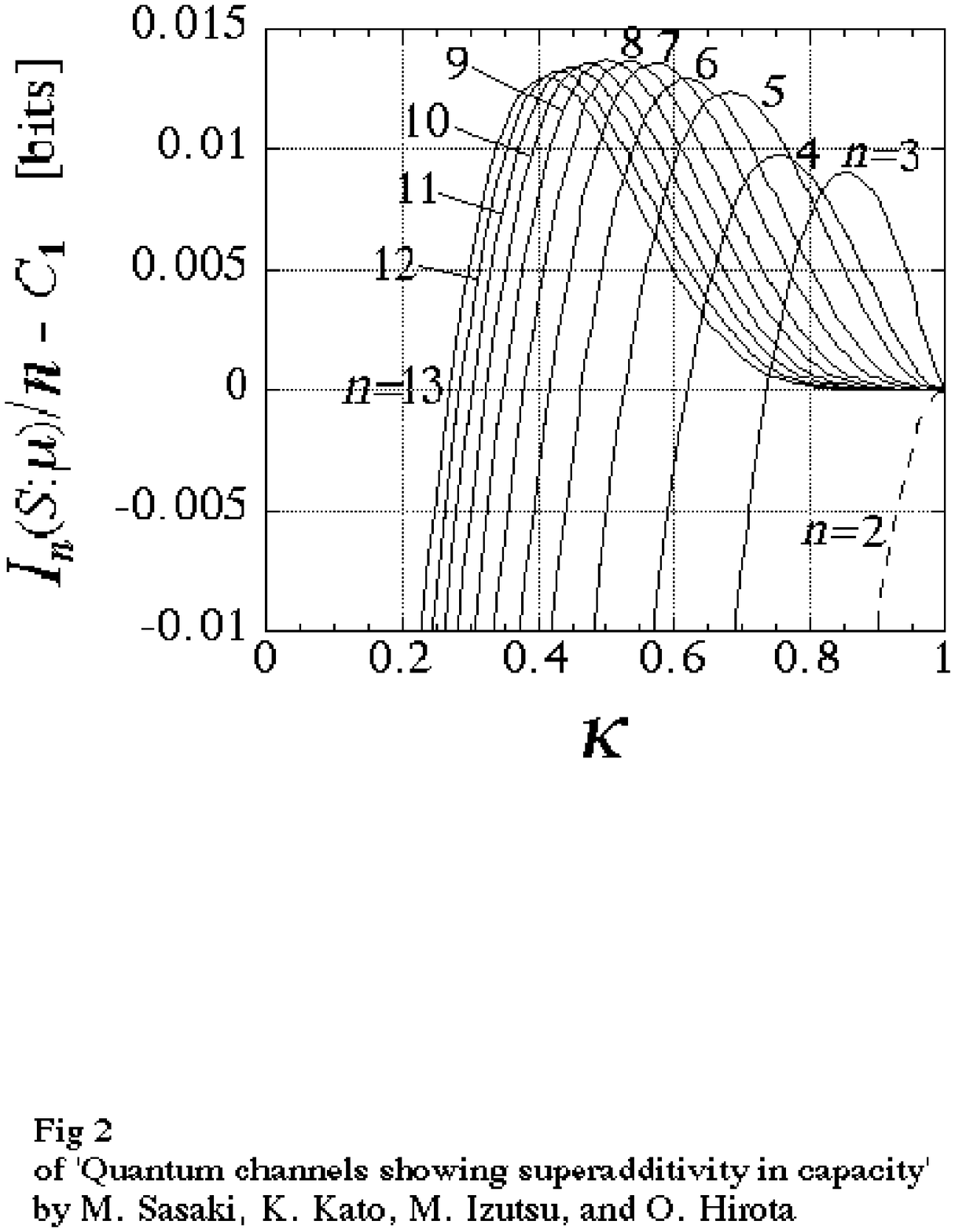}}
\caption{The difference between the mutual information per letter state 
$I_n(S:\mu)/n$ and $C_1(\hat{\bf s})$ for $n=2\sim13$}.  The dashed line
corresponds to the case of $n=2$, while the solid lines correspond to the
case of $n=3\sim13$.
\label{fig2}
\end{figure}

\begin{figure}
\centerline{\epsffile{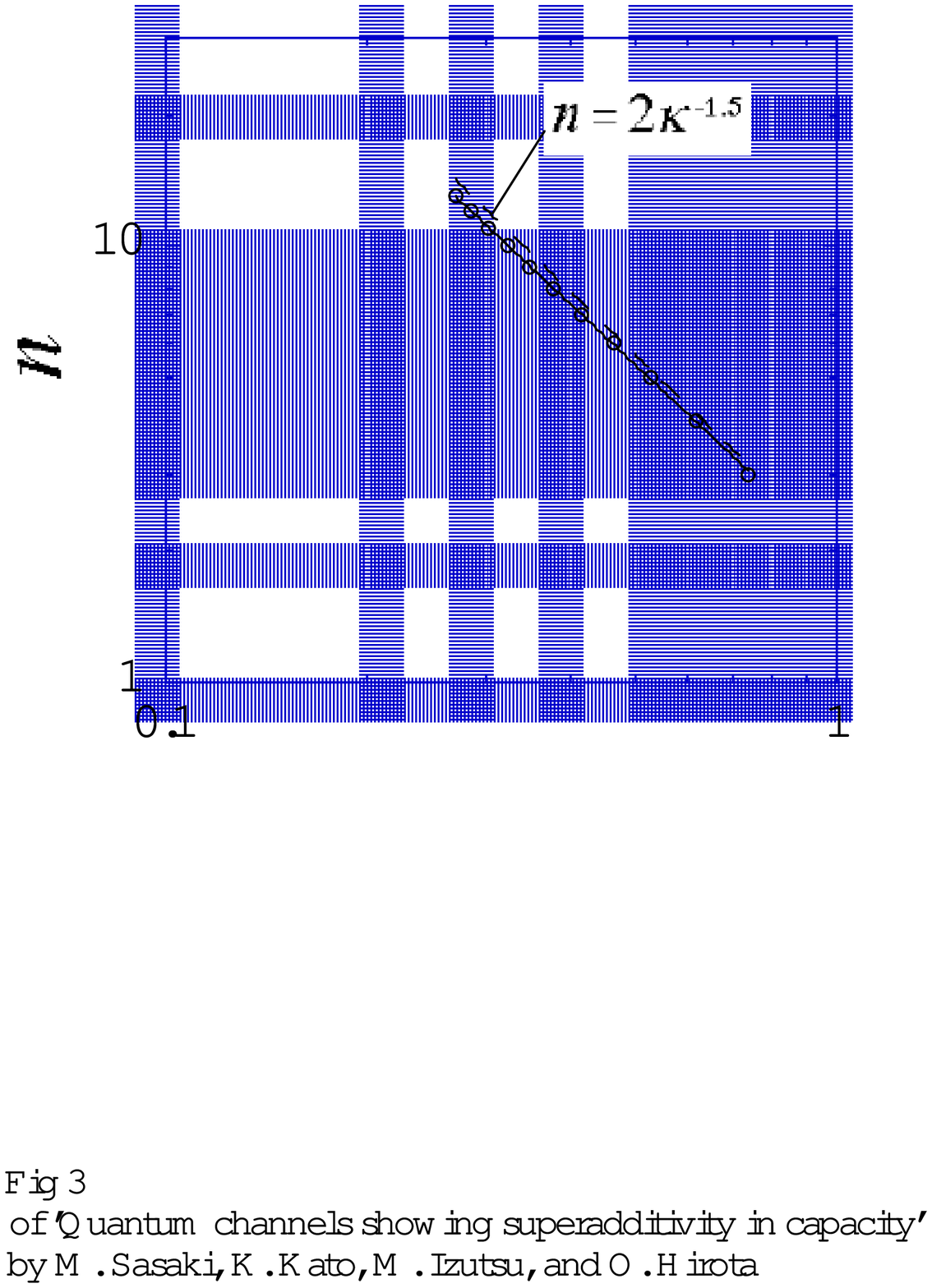}}
\caption{The relation between $n$ and $\kappa_\ast$ $(<1)$ at which
$I_n(S:\mu)/n=C_1(\hat{\bf s})$ holds. $\kappa_\ast$'s are denoted by the
circles. The solid line is just a guide for eye. The dashed line
corresponding to the curve of $n=2\kappa^{-1.5}$.}
\label{fig3}
\end{figure}

\begin{figure}
\centerline{\epsffile{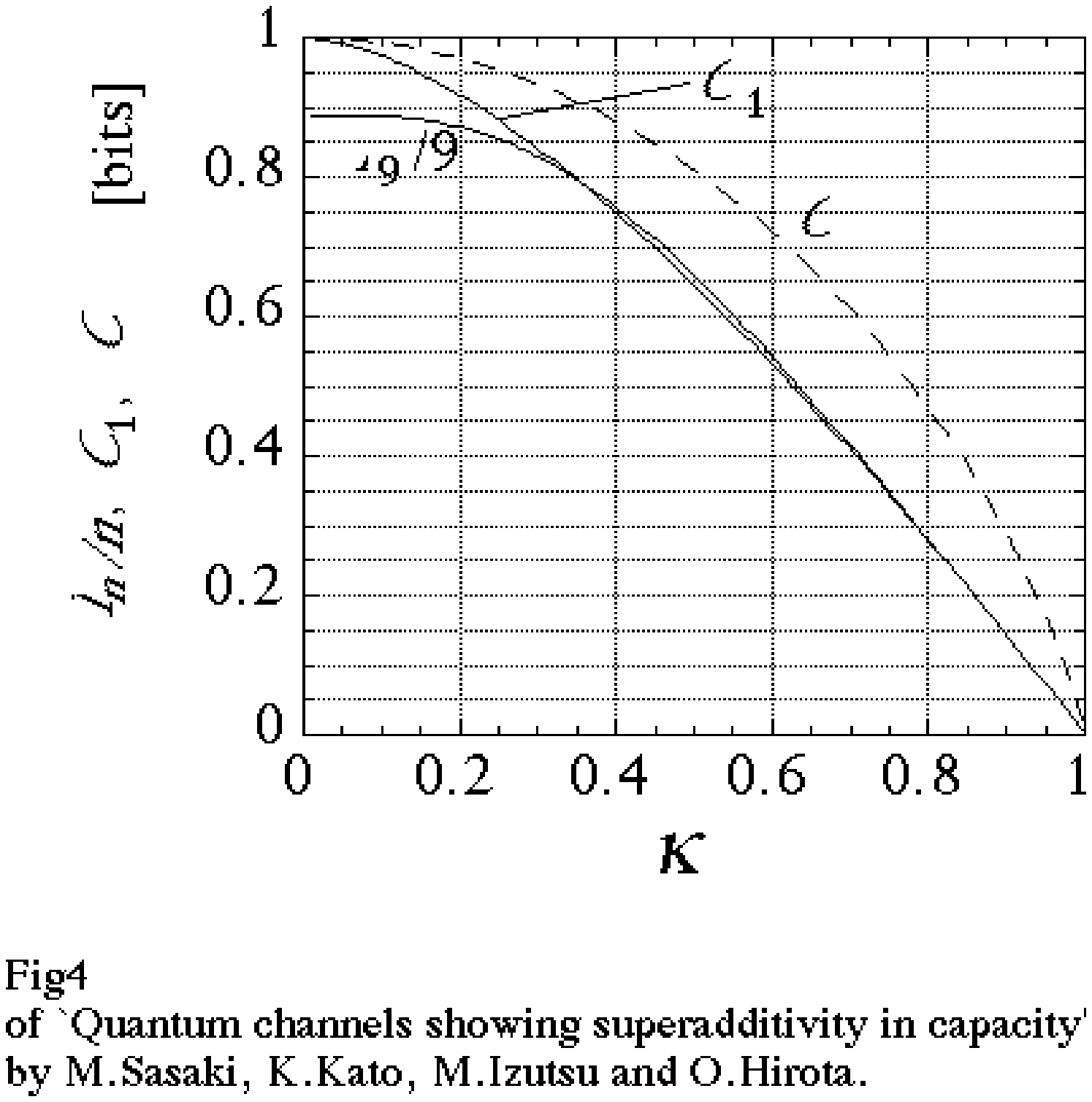}}
\caption{$C(\hat{\bf s})$, $I_9(S:\mu)/9$ and $C_1(\hat{\bf s})$ as
functions of $\kappa$. }
\label{fig4}
\end{figure}

\begin{figure}
\centerline{\epsffile{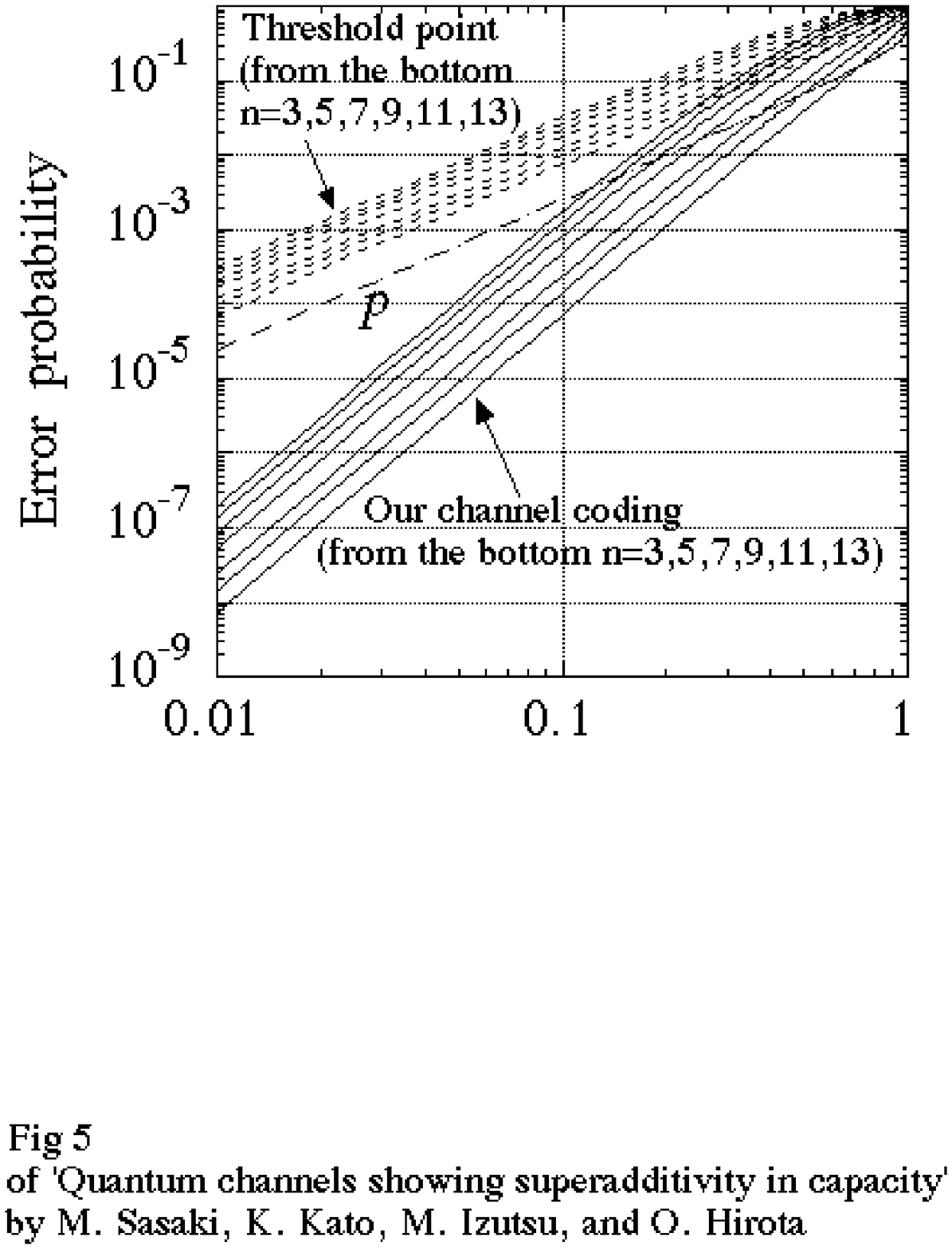}}
\caption{The minimum average error probabilities corresponding to the code
$[[n,n-1,2]]$ (solid lines), the initial channel, i.e., $p$ (dashed line),
and the {\it threshold points} (dotted lines) as functions of $\kappa$.}
\label{fig5}
\end{figure}

\begin{figure}
\centerline{\epsffile{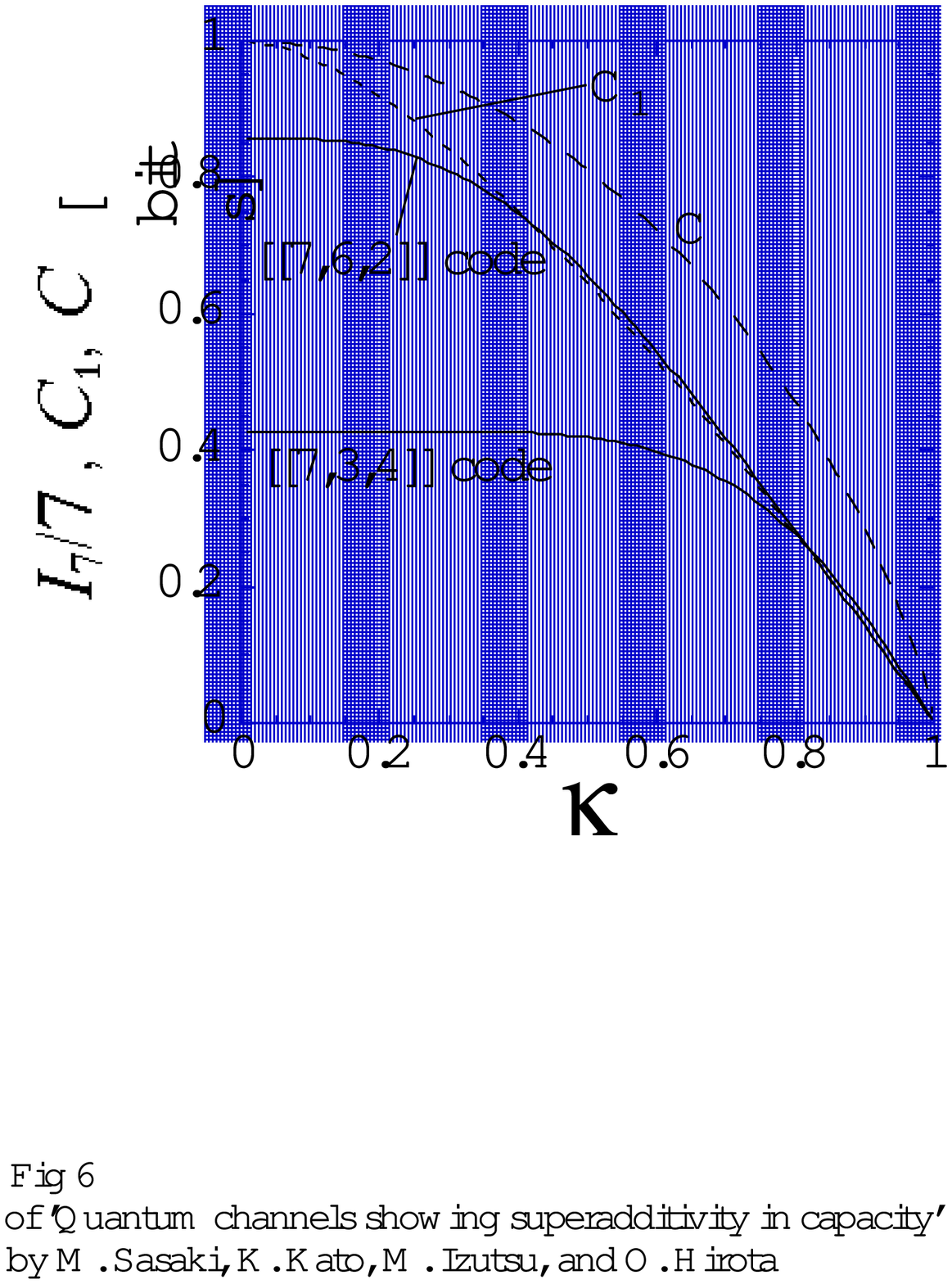}}
\caption{$C(\hat{\bf s})$, $I_7(S:\mu)/7$ for both of the [[7,3,4]] simplex
code and the [[7,6,2]]  code,  and $C_1(\hat{\bf s})$ as functions of
$\kappa$.}
\label{fig6}
\end{figure}

\begin{figure}
\centerline{\epsffile{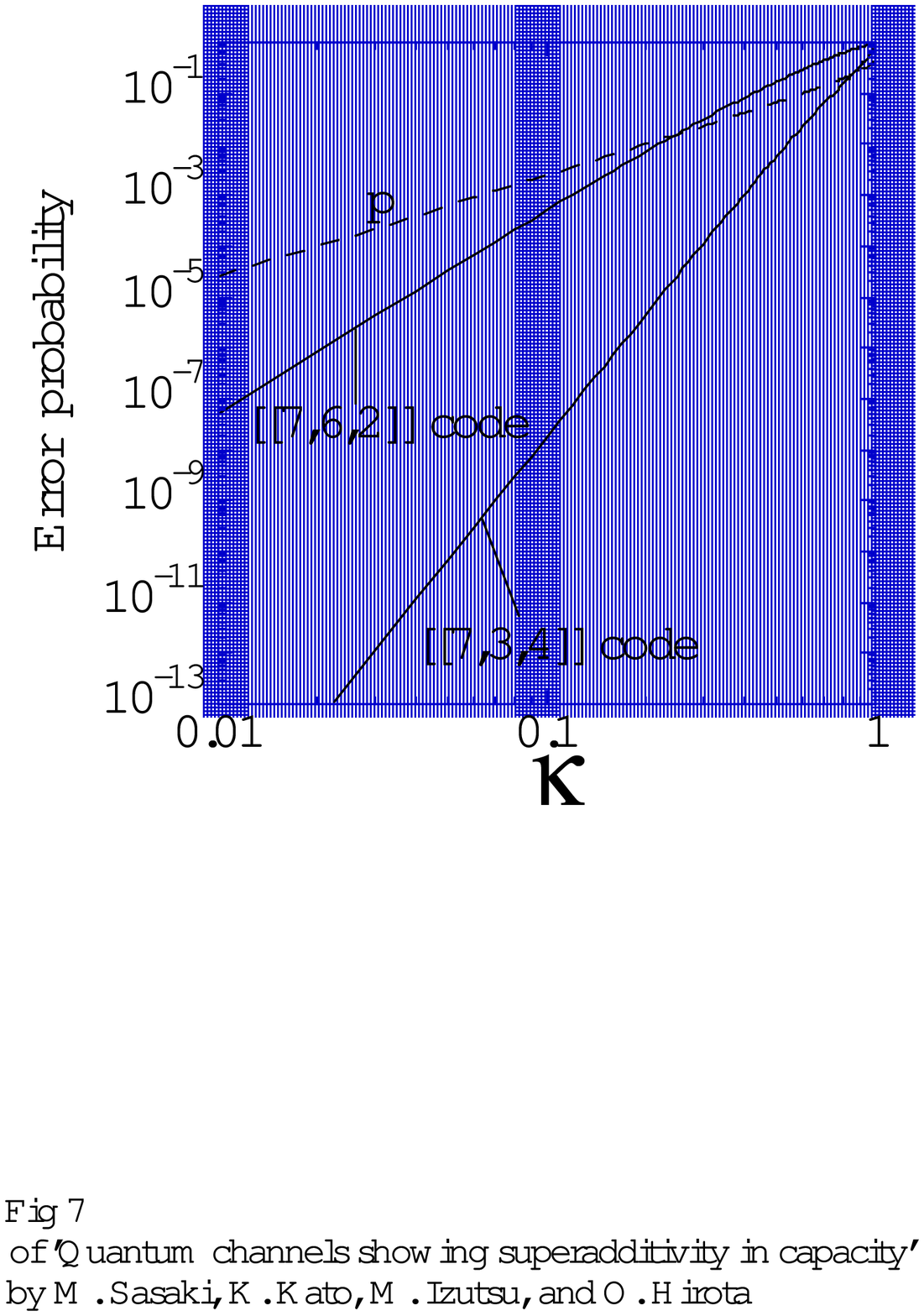}}
\caption{The minimum average error probabilities corresponding to the
[[7,3,4]] simplex code, the [[7,6,2]]  code ,  and the initial channel,
i.e., $p$ (dashed line) as functions of $\kappa$. }
\label{fig7}
\end{figure}

\begin{figure}
\centerline{\epsffile{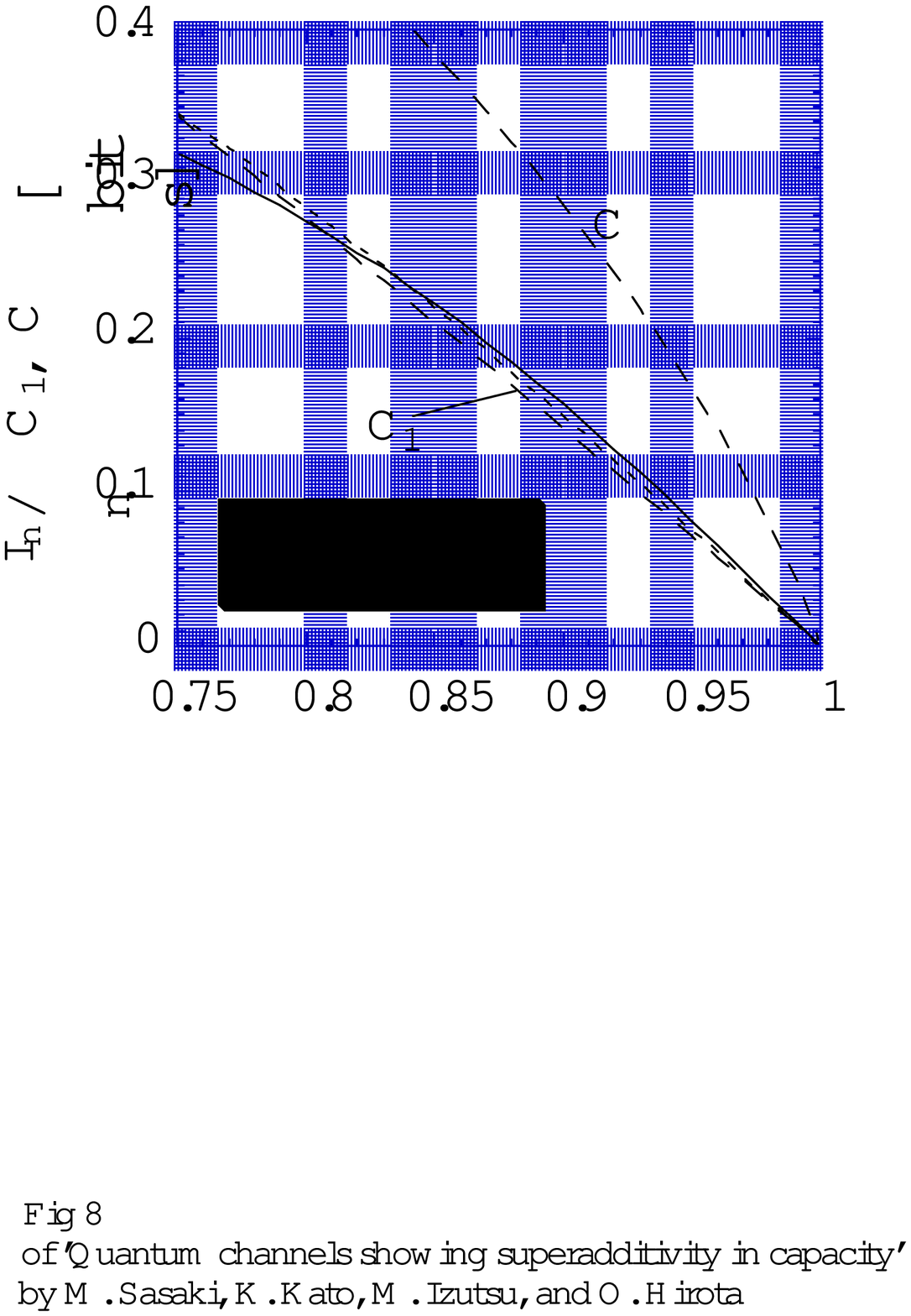}}
\caption{The mutual informations per letter corresponding to the [[7,3,4]]
simplex code (solid line) and the code [[3,2,2]] (dashed line) as functions
of $\kappa$. }
\label{fig8}
\end{figure}


\begin{references}

\bibitem {Holevo73_bound} A. S. Holevo, Probl. Peredachi Inform. vol 9, no.
3, 3111 (1973). 
\bibitem {Holevo79_QuantCap} A. S. Holevo, Probl. Peredachi Inform. vol 15,
no. 4, 3 (1979). 
\bibitem {Hausladen96_coding} P. Hausladen, R. Jozsa, B. Schumacher, M.
Westmoreland and W. K. Wootters,  Phys. Rev. A{\bf 54}, 1869 (1996). 
\bibitem {Holevo96_coding1} A. S. Holevo, Report No. quant-ph/9611023, Nov.
to appear in IEEE Trans. {\bf IT}. 
\bibitem {Schumacher97_coding} B. Schumacher and  M. Westmoreland,  Phys.
Rev. A{\bf 56}, 131 (1997). 
\bibitem {Holevo97_coding2} A. S. Holevo, Report No. quant-ph/9708046, 27
Aug. (1997). 
\bibitem{Helstrom_QDET} C. W. Helstrom : {\it Quantum Detection and
Estimation Theory} (Academic Press, New York,  1976).
\bibitem {Holevo_SubOptMeas78} A. S. Holevo, Theory Prob. Appl., vol. 23,
411, June(1978). 
\bibitem {Hausladen_SubOptMeas95} P. Hausladen and W. K. Wootters, J. Mod.
Opt. {\bf 41}, 2385 (1994). 
\bibitem {FuchsPeres96} C. A. Fuchs and A. Peres, Phys. Rev. A{\bf 53},
2038 (1996)
\bibitem {Ban97_C1} M. Ban, K. Yamazaki, and O. Hirota, Phys. Rev. A{\bf
55}, 22 (1997). 
\bibitem {Osaki97_C1} M. Osaki, M. Ban, and O. Hirota, to appear in J. Mod.
Opt. (1997). 
\bibitem {Holevo73_condition} A. S. Holevo, J. Multivar. Anal. {\bf 3},
337, (1973).
\bibitem {Osaki} M. Osaki and O. Hirota, {\it Quantum Communication and
Measurement}, pp401-409,
( ed. by Belavkin, Hirota, and Hudson, Plenum Publishing, New York, 1995);
M. Osaki, M. Ban, and O. Hirota, Phys. Rev. A{\bf 54},  1691, (1996).
\bibitem {Helstrom82} C. W. Helstrom, IEEE Trans., {\bf IT-28}, 359, (1982).
\bibitem {BanKurokawa_SqRt} M. Ban, K. Kurokawa, R. Momose, and O. Hirota,
Inter. J. Theor. Phys. {\bf 36}, 1269 (1997). 
\bibitem {Hirota88_RQSC} O. Hirota, Opt. Commun., {\bf 67}, 204, (1988). 
\bibitem {Sasaki's} M. Sasaki and O. Hirota: Phys. Lett. {\bf A210}, 21
(1996); ibid{\bf A224}, 213 (1997); M. Sasaki, T. S. Usuda, and O. Hirota,
A. S. Holevo, Phys. Rev. A{\bf 53},  1273, (1996); M. Sasaki and O. Hirota,
ibid{\bf 54}, 2728, (1996). 
\bibitem {Barenco95} A. Barenco, C. H. Bennett, R. Cleve, D. P. M.
DiVincenzo, N. Margolus, P. Shor, T. Sleator, J. A. Smolin, and H.
Weinfurter, Phys. Rev. A{\bf 52},  3457, (1995).        
\bibitem {Reck94} M. Reck, A. Zeilinger, H. J. Bernstein, and P. Bertani,
Phys. Rev. Lett. {\bf 73}, 58 (1994).        
\bibitem {Sleator95} T. Sleator and H. Weinfurter, Phys. Rev. Lett. {\bf
74}, 4087 (1995).
\bibitem {Turchette95}   Q. A. Turchette, C. J. Hood, W. Lange, H. Mabuchi,
and H. J. Kimble,    Phys. Rev. Lett. {\bf 75}, 4710 (1995).  
\bibitem{Holevo97_private_commun}  The authors are indebted to a private
communication from A. S. Holevo for the proofs of Theorem 2 and 3.         
             
\bibitem {Sasaki97_SupAdd} M. Sasaki, K. Kato, M. Izutsu, and O. Hirota, to
appear  in Phys. Lett. A (1997). 



\end{references}
\end{document}